\documentclass[11pt]{article}
\usepackage[left=1in,right=1in]{geometry}
\usepackage{empheq}
\usepackage{amssymb}
\usepackage{amsmath}
\usepackage{dsfont}
\usepackage{graphicx}
\usepackage[pdfstartview=FitH,pdfpagemode=None]{hyperref}
\usepackage{color}
\usepackage{slashed}
\usepackage{multirow}

\makeatletter\@addtoreset{equation}{section}\makeatother

\newcommand{\be}{\begin{equation}}
\newcommand{\ee}{\end{equation}}
\newcommand{\ba}{\begin{eqnarray}}
\newcommand{\ea}{\end{eqnarray}}
\newcommand{\bea}{\begin{eqnarray}}
\newcommand{\eea}{\end{eqnarray}}
\newcommand{\nin}{\noindent}
\def\s{\sigma}
\newcommand{\vev}[1]{{\left< {#1} \right>}}

\begin{document}
\begin{titlepage}
\hfill MCTP-16-03

\begin{center}
{\Large \bf Toward precision holography with supersymmetric Wilson loops}
\vskip .5 cm

Alberto Faraggi,${}^{a}$ Leopoldo A. Pando Zayas,${}^{b}$ Guillermo A. Silva,${}^{c}$ and Diego Trancanelli${}^{d}$
\vskip .5 cm
{\it ${}^{a}$ Instituto de F\'\i sica, Pontificia Universidad Cat\'olica de Chile}\\
{\it  Casilla 306, Santiago, Chile}
\vskip .3 cm
{\it ${}^{b}$ The Abdus Salam International Centre for Theoretical Physics}\\
{\it Strada Costiera 11, 34014 Trieste, Italy}\\
\vspace{.3cm}
{\it ${}^b$  Michigan Center for Theoretical Physics,  Department of Physics}\\
{\it University of Michigan, Ann Arbor, MI 48109, USA}\\
\vskip .3 cm
{\it ${}^c$ Instituto de F\'\i sica de La Plata - CONICET \& Departamento de F\'\i sica - UNLP}\\
{\it C.C. 67, 1900 La Plata, Argentina}
\vskip .3 cm
{\it ${}^d$ Institute of Physics, University of S\~ao Paulo,}
{\it 05314-970  S\~ao Paulo, Brazil}
\end{center}
\vskip 1 cm

\begin{abstract}
\nin
We consider certain 1/4 BPS Wilson loop operators in $SU(N)$ ${\cal N}=4$ supersymmetric Yang-Mills theory, whose expectation value can be computed exactly via supersymmetric localization. Holographically, these operators are mapped to fundamental strings in $AdS_5\times S^5$. The string on-shell action reproduces the large $N$ and large coupling limit of the gauge theory expectation value and, according to the AdS/CFT correspondence, there should also be a precise match between subleading corrections to these limits. We perform a test of such match at next-to-leading order in string theory, by deriving the spectrum of quantum fluctuations around the classical string solution and by computing the corresponding 1-loop effective action. We discuss in detail the supermultiplet structure of the fluctuations. To remove a possible source of ambiguity in the ghost zero mode measure, we compare the 1/4~BPS configuration with the 1/2 BPS one, dual to a circular Wilson loop. We find a discrepancy between the string theory result and the gauge theory prediction, confirming a previous result in the literature. We are able to track the modes from which this discrepancy originates, as well as the modes that by themselves would give the expected result. 
\end{abstract}
\end{titlepage}

\tableofcontents

\section{Introduction}

The AdS/CFT correspondence provides a paradigm wherein a field theory is equivalent to a string theory containing gravity \cite{Maldacena:1997re}. The most studied and best understood example of this correspondence conjectures the equivalence of $SU(N)$ ${\cal N}=4 $ super Yang-Mills theory and type IIB string theory on $AdS_5\times S^5$ with $N$ units of Ramond-Ramond (RR) five-form flux. There are various levels at which this correspondence can be tested. The `weakest' level is the limit of large $N$ and strong 't Hooft coupling on the field theory side, whose dual string theory is well described by classical supergravity. Going beyond this limit is, potentially, a conceptually fruitful endeavor.  An ideal arena were this can be achieved is the study of non-local supersymmetric operators such as the Wilson loops.

Very soon after the Maldacena correspondence was put forward, it was proposed that the vacuum expectation value of the 1/2 BPS circular Wilson loop, arguably the simplest non-local supersymmetric operator, is captured by a Gaussian matrix model \cite{Erickson:2000af,Drukker:2000rr}. This conjecture was later proven by Pestun \cite{Pestun:2007rz}, using the technique of supersymmetric localization. For the case of the fundamental representation of $SU(N)$, the vacuum expectation value of this operator is known exactly for any $N$ and any 't Hooft coupling $\lambda=g^2_\textrm{YM} N$ in terms of generalized Laguerre polynomials~\cite{Drukker:2000rr}:
\bea
    \langle W\rangle_{\textrm{circle}}&=&\frac{1}{N}L^1_{N-1}\left(-\frac{\lambda}{4N}\right)e^{\lambda/8N}     \cr
    &\simeq & \frac{2}{\sqrt{\lambda}}I_1(\sqrt{\lambda})+\frac{\lambda}{38 N^2}I_2(\sqrt{\lambda})+\frac{\lambda^2}{1280 N^4}I_4(\sqrt{\lambda})+ \ldots \cr
    &\simeq & \exp\left(\sqrt{\lambda} -\frac{3}{4}\ln \lambda -\frac{1}{2}\ln \frac{\pi}{2} +\ldots \right).
    \label{halfBPSoperator}
\eea
The first line is exact in $N$ and $\lambda$, the second line is an expansion in large $N$, and in the last line the large $\lambda$ limit is also taken.

Having an exact field theory answer poses one of the simplest, yet elusive, tests of the AdS/CFT correspondence. The situation is akin to a high precision test of the AdS/CFT correspondence, where the field theory side provides the ``experimental'' side and string theory is the theory that should match the experimental results. Indeed, there has been a fairly concerted effort in trying to match the field theory answer (\ref{halfBPSoperator}) with the 1-loop corrected answer coming from holography. The first efforts date back over a decade and a half \cite{Drukker:2000ep}. More recently, the 1-loop correction has been revisited using different methods in \cite{Kruczenski:2008zk} and \cite{Buchbinder:2014nia}, leading to
\be
\langle W\rangle_{\textrm{circle}}=\exp\left(\sqrt{\lambda} -\frac{1}{2}\ln (2\pi)+ \ldots\right).
\ee
The main missing term in this formula is the $-(3/4) \ln \lambda$. There is also a numerical discrepancy in the constant term. This discrepancy has been attributed to ghost zero modes in the corresponding string amplitude \cite{Drukker:2000ep,Kruczenski:2008zk,Buchbinder:2014nia}. There are also similar discrepancies when confronting field theory results with holographic computations at 1-loop level for Wilson loops in higher rank representations  as summarized in \cite{Faraggi:2014tna}, albeit in those cases the functional dependence matches.

Our driving motivation is not a hidden suspicion of the validity of the AdS/CFT correspondence, rather we believe that, by carefully considering such discrepancies, we might learn something about the intricacies of computing string theory on curved backgrounds with RR fluxes, thus broadening  the class of problems which the AdS/CFT can tackle at the quantum level. In this sense our philosophy is summarized in the following question: {\it What can we learn about string theory in curved backgrounds from having exact results on the dual, gauge theory side?} 

With this general motivation in mind, we turn to the study of certain $1/4$ BPS Wilson loop introduced in \cite{Drukker:2005cu,Drukker:2006ga} and further studied in \cite{Drukker:2006zk,Drukker:2007dw,Drukker:2007yx,Drukker:2007qr}. These loops are called ``latitude'' Wilson loops and from the field theory point of view are quite similar to the $1/2$ BPS circle. The latitudes are defined in terms of a parameter, $\theta_0\in [0,\pi/2]$, which selects a  latitude on an $S^2$ on which the loop is supported, see the next section for more details. The vacuum expectation value of this operator is conjectured to be given by a simple re-scaling of the 't Hooft coupling in the exact expression for the $1/2$ BPS Wilson loop \cite{Drukker:2006ga,Drukker:2007yx,Drukker:2007qr}:
\bea
\langle W\rangle_{\textrm{latitude}}&=&\frac{1}{N}L^1_{N-1}\left(-\frac{\lambda'}{4N}\right)e^{\lambda'/8N}   \,,
\eea
where $\lambda'=\lambda \cos^2\theta_0$. In fact, this conjecture extends to a larger class of (generically 1/8 BPS) Wilson loops, the so-called DGRT loops, defined as generic contours on an $S^2$ \cite{Drukker:2007dw,Drukker:2007yx,Drukker:2007qr}, of which the latitude is a special example with enhanced supersymmetry. This conjecture has passed several non-trivial tests. In perturbation theory, it has been checked explicitly for specific examples of DGRT loops, and correlators thereof, up to third order, see for example \cite{Bassetto:2008yf,Bassetto:2009rt,Bassetto:2009ms,Giombi:2009ms}. At strong coupling, it has been checked in \cite{Drukker:2006ga,Drukker:2007qr} by constructing the corresponding string configurations and evaluating their on-shell action. Finally, localization has been applied in \cite{Pestun:2009nn}, where it was shown\footnote{The proof of localization is somewhat incomplete, since it lacks a computation of the 1-loop determinants.} that these loops reduce to the Wilson loops in the zero-instanton sector of (purely bosonic) Yang-Mills theory on a two-sphere, which is an exactly solvable theory \cite{Migdal:1975zg}, see for example \cite{Witten:1992xu,Bassetto:1998sr}. 

Holographically, the 1/4 BPS latitude gets mapped to a macroscopic string in $AdS_5\times S^5$, which not only extends on the $AdS_5$ part of the geometry, as the 1/2 BPS string does, but it also wraps a cup in the $S^5$ part. For some recent investigations into these configurations see, for example,  \cite{Miwa:2015bta} and \cite{Liu:2015qsa}. The main idea of this paper is  to compute the 1-loop effective action for this string and compare it with the effective action for the 1/2 BPS string. Since both strings have a world-sheet with the topology of a disk, the expectation is that the issues related to the ghost zero modes, which we have mentioned above, might cancel. More specifically, we consider the ratio
\be
\frac{\vev{W}_\textrm{latitude}}{\vev{W}_\textrm{circle}}
\simeq\exp\left(
\sqrt{\lambda}(\cos\theta_0-1)-\frac{3}{2}\ln\cos\theta_0+\ldots
\right)\,,
\label{prediction}
\ee
with the intent of recovering the $-(3/2) \ln\cos\theta_0$ term from the string theory 1-loop effective action.

The paper is organized as follows. We review various field theoretic aspects of the 1/4 BPS Wilson loop in Sec.~\ref{Sec:FT} and the classical string solution in Sec.~\ref{Sec:Classical}. We present a derivation and analysis of the fluctuations in Sec.~\ref{Sec:Fluctuations}. In particular, we show how they are neatly organized in representations of the supergroup $SU(2|2)$. We compute the determinants in Sec.~\ref{Sec:Determinants} and the 1-loop effective action in Sec.~\ref{sec:1loopaction}. We finally conclude with some comments and outlook in Sec.~\ref{Sec:Conclusions}. We relegate a number of explicit technical calculations to the appendices.

{\bf Note 1:} As we were in an advanced stage of this project (partial progress having been reported in \cite{talks}), the paper \cite{Forini:2015bgo} appeared. There is certainly a lot of overlap. Although conceptually similar, our work has some technical differences with \cite{Forini:2015bgo}, which we highlight. In particular, we stress the role of group theory in the spectrum of fluctuations and in the sums over energies, we have a different treatment of the fermionic spectral problem, for we consider the linear operator instead of the quadratic one, and we use different boundary conditions for the fermions. Moreover, our treatment of the 1-loop effective action is fully analytical, whereas \cite{Forini:2015bgo} resorted to numerics.

{\bf Note 2:} In this revision, we correct a critical mistake in the original manuscript submitted to the arXiv that  alters our conclusions. Instead of the agreement between gauge theory expectation and string theory claimed in the v1, we do find a finite discrepancy, precisely equal to the remnant reported in \cite{Forini:2015bgo}. One advantage of having an analytical treatment, as we do here,  is that we are able to track the origin both of the expected result ({\it i.e.}, the $-(3/2)\ln\cos\theta_0$ term) and of the discrepancy to certain specific modes. We hope this might be useful for future investigations, as we comment in the conclusions.

\section{The $1/4$ BPS latitude in ${\cal N}=4$ super Yang-Mills}
\label{Sec:FT}

We start with a brief review of the gauge theory side \cite{Drukker:2006ga,Drukker:2007qr}. The 1/4 BPS latitude Wilson loop (in the fundamental representation of $SU(N)$) is defined as
\begin{empheq}{align}
    W(C)=\frac{1}{N}\textrm{Tr}\,{\cal P}\exp\int_C ds\left(iA_{\mu}\dot{x}^{\mu}+|\dot x|\Phi_I \, n^I(s)\right)\,,
\end{empheq}
where ${\cal P}$ denotes path ordering along the loop and $C$ labels a curve parametrized as
\bea
x^\mu(s)=(\cos s,  \sin s, 0,0)\,,\qquad
n^I(s)=(\sin\theta_0\cos s, \sin\theta_0\sin s, \cos\theta_0, 0,0,0).
\eea
This operator interpolates between the 1/2 BPS circle, corresponding to $\theta_0=0$, and the so-called Zarembo loops \cite{Zarembo:2002an} at $\theta_0=\pi/2$. It preserves a $SU(2|2)$ subgroup of the superconformal group $SU(2,2|4)$ of ${\cal N}=4$ super Yang-Mills,  for more detail see App. B.2 of \cite{Drukker:2007qr}. The bosonic symmetries are given by
\be
SU(2) \times U(1) \times SU(2)_B\,.
\label{symmetries}
\ee
The first $SU(2)$ factor is a remnant of the conformal group, broken by the presence of the latitude circle. This is, in fact, the same $SU(2)$ factor from $SO(4,2)$ which is also preserved by the 1/2 BPS circle, although the symmetry is realized differently in the two cases. Note, in passing, that the 1/4 BPS loop does not preserve the $SL(2,\mathbb{R})$ subgroup of $SO(4,2)$ preserved by the 1/2 BPS circle. In the holographic dual, this will manifest itself in the fact that the induced metric on the string world-sheet is not $AdS_2$, as it is the case for the string corresponding to the 1/2 BPS circle. The  $U(1)$ symmetry in (\ref{symmetries}) mixes Lorentz and R-symmetry transformations
\be
 C = J_{12} + J^A_{12}\,,
\ee
 with $J_{12}$ coming from $SL(2,\mathbb{R})$ and $J^A_{12}$ from the $SU(2)_A$ subgroup of  the $SU(4)$ R-symmetry. In the holographic dual, this symmetry is implemented as translations along the $\psi$ and $\phi$ coordinates, as we shall see presently. The last $SU(2)$ is the $SU(2)_B$ subgroup of the R-symmetry. This can be understood by noticing that the loop is only defined in  terms of the scalars $\Phi_{1,2,3}$, which are rotated by $SU(2)_A$, whereas the other three scalar fields $\Phi_{4,5,6}$, which do not appear in the Wilson loop, are rotated by $SU(2)_B$. From the holographic point of view, as we will review in the upcoming section, one can think of this symmetry  in terms of the embedding coordinates of the sphere where an $SO(3)$ is explicit.

\section{Review of the classical string solution}
\label{Sec:Classical}

In this section we review the classical string solution dual to the 1/4 BPS latitude Wilson loop \cite{Drukker:2006ga,Drukker:2007qr}. The supergravity background is given by $AdS_5\times S^5$ with a five-form  RR flux and the $AdS_5$ metric conveniently expressed as a foliation over $\mathds{H}_2\times\mathds{H}_2$
\begin{empheq}{align}
	 ds^2_{AdS_5}&=\cos^2 u\left(d\rho^2+\sinh^2\rho\,d\psi^2\right)+\sin^2 u \left(d\vartheta^2+\sinh^2\vartheta\,d\varphi^2\right)-du^2\,.
\label{adsfol}
\end{empheq}
We have set the radius equal to 1. The Euclidean continuation is achieved by taking $u\rightarrow iu$ and $\vartheta\rightarrow i\vartheta$, such that the $EAdS_5$ metric becomes now a foliation over $\mathds{H}_2\times S^2$
\begin{empheq}{align}\label{Eq:AdS5}
	 ds^2_{AdS_5}&=\cosh^2 u\left(d\rho^2+\sinh^2\rho\,d\psi^2\right)+\sinh^2 u\left(d\vartheta^2+\sin^2\vartheta\,d\varphi^2\right)+du^2\,.
\end{empheq}
The metric on $S^5$ is taken to be
\begin{empheq}{align}\label{Eq:S5}
	d\Omega_5^2&=d\theta^2+\sin^2\theta\,d\phi^2+\cos^2\theta\left(d\xi^2+\cos^2\xi\,d\alpha_1^2+\sin^2\xi\,d\alpha_2^2\right)\,,
\end{empheq}
and the 4-form potential reads
\begin{empheq}{alignat=5}
	C_{(4)}&=\left(\frac{1}{8}\sinh(4u)-\frac{u}{2}\right)\,\textrm{vol}\left(AdS_2\right)\wedge\textrm{vol}\left(S^2\right)\,,
\end{empheq}
with corresponding field strength $F_{(5)}=-4\left(1+*\right)\textrm{vol}\left(AdS_5\right)$.

The string has world-sheet coordinates $(\tau,\sigma)$ and its embedding in the background above is given by \cite{Drukker:2006ga}:
\begin{empheq}{align}
	\sinh\rho=\frac{1}{\sinh\sigma}\,, \qquad \psi=\tau\,,
	\qquad
	u=0\,,\qquad 
	\sin\theta=\frac{1}{\cosh\left(\sigma_0+\sigma\right)}\,,\qquad 	\phi=\tau\,, 
	\label{Eq:Solution}
\end{empheq}
where $\sigma_0$ sets the range of  values of $\theta$, namely, $0\le \theta\le \theta_0$, with
\begin{empheq}{align}
	\sin\theta_0&=\frac{1}{\cosh\sigma_0}\,.
	\label{sigma0theta0}
\end{empheq}
The remaining coordinates take arbitrary constant values. The string world-sheet forms a cap through the north pole of the $S^5$. The sign of $\sigma_0$ determines whether the world-sheet starts above ($\sigma_0>0$) or below the equator ($\sigma_0<0$),  this last case being unstable under fluctuations \cite{Drukker:2006ga}.

The induced geometry on the string world-sheet is
\begin{empheq}{align}
	ds^2&=\left(\sinh^2\rho+\sin^2\theta\right)d\tau^2+(\rho'^2+\theta'^2)d\sigma^2\,.
\end{empheq}
Since the solution satisfies $\rho'=-\sinh\rho$ and $\theta'=-\sin\theta$, we can write the induced metric  as 
\begin{empheq}{align}
	ds^2&=\left(\sinh^2\rho+\sin^2\theta\right)\left(d\tau^2+d\sigma^2\right)\,.
\end{empheq}
In the following, we shall denote the overall conformal factor as
\begin{empheq}{alignat=5}
	A\equiv\sinh^2\rho+\sin^2\theta
	=\frac{1}{\sinh^2\sigma} +\frac{1}{\cosh^2(\sigma_0+\sigma)}\,,
	\label{A}
\end{empheq}
where in the last equality we have used the explicit solution for the embeddings $\rho(\sigma)$ and $\theta(\sigma)$ in~(\ref{Eq:Solution}). 

In the $\sigma_0\rightarrow\infty$  limit, the range of $\theta$ shrinks to a point. In this sense the $1/4$ BPS solution reduces to the $1/2$ BPS one, where $\theta$ is but  a point on $S^5$ and the string world-sheet has an $AdS_2$ geometry. This has the topology of a disc plus a point. The disk along the $AdS_2$ part has radial coordinate $\sigma\in[0,\infty)$ (with boundary located at $\sigma=0$) and angular coordinate $\tau\sim\tau+2\pi$. The cap on $S^2$ is contractible and, consequently, equivalent to the point on the north pole which corresponds to the solution in the $1/2$ BPS case.

The string action can be evaluated on-shell on this classical solution. The result, after an appropriate renormalization, is  \cite{Drukker:2006ga}
\bea
S^{(0)}=-\sqrt{\lambda}\cos\theta_0\,.
\eea
Since $\langle W \rangle\simeq  \exp\left(-S^{(0)}\right)=\exp\left(\sqrt{\lambda}\cos\theta_0\right)$, we recover, at the classical level, the expectation (\ref{prediction}) from field theory.  

\subsection{Symmetries of the classical solution}

In \cite{Drukker:2007qr} it was shown that the 1/4 BPS latitude preserves an $SU(2|2)$ subgroup of the superconformal group of ${\cal N}=4$ super Yang-Mills. The corresponding  bosonic subgroup is  $SU(2)\times U(1)\times SU(2)_B\simeq SO(3)\times SO(2)\times SO(3)$.

One of the simplest way to see how the embedding preserves $SO(3)\times SO(3)$ is by expressing the solution in the embedding coordinates $X_i$.  For $AdS_5$ we have $-X_0^2+X_1^2+X_2^2+X_3^2+X_4^2+X_5^2=-1$, with the solution taking the form
\be
X_0=\text{coth}\s, \quad X_1=\text{cosech}\s \, \cos\tau, \quad X_2=\text{cosech}\s \, \sin\tau, \quad X_3=X_4=X_5=0.
\ee
One explicitly sees that there is an $SO(3)$ group that rotates the coordinates  $(X_3,X_4,X_5)$ without affecting the solution.
On the $S^5$ side, whose equation we write as $Y_1^2+Y_2^2+Y_3^2+Y_4^2+Y_5^2+Y_6^2=1$, we have
\be
Y_1=\text{sech}(\s_0+ \s)\cos\tau, \quad Y_2=\text{sech}(\s_0+ \s)\sin\tau, \quad Y_3=\tanh(\s_0+ \s), \quad Y_4=Y_5=Y_6=0\,,
\ee
where $\tanh\s_0=\cos\theta_0$. Similarly,  there is an $SO(3)$ group that rotates the coordinates  $(Y_4,Y_5,Y_6)$ without affecting the solution. There is an $SO(2)$ rotation realized in the plane $(X_1,X_2)$ and an $SO(2)$ rotation realized in the plane $(Y_1,Y_2)$. These symmetries are identified as translations in $\tau$, as can be clearly seen in the classical solution $\psi=\tau=\phi$ in (\ref{Eq:Solution}). 

We shall show later on that the string fluctuations around the 1/4 BPS solutions are neatly organized in multiplets of this $SU(2|2)$ supergroup.

\section{Quadratic fluctuations}
\label{Sec:Fluctuations}

Having reviewed the classical solution dual to the 1/4 BPS latitude Wilson loop and its symmetries, in this section we derive the corresponding spectrum of excitations. For the case of the 1/2 BPS circular Wilson loop, the dual solution and its perturbations have been known for quite some time, see for example \cite{Drukker:2000ep,Kruczenski:2008zk,Buchbinder:2014nia}. Similar studies for holographic duals of Wilson loops in higher representations  include \cite{Camino:2001at,Faraggi:2011ge,faraggi:2011bb}.

We will start by giving a general expression for the quadratic fluctuations of the type IIB string in $AdS_5\times S^5$ and then specialize to the case of the 1/4 BPS string dual to the latitude Wilson loop. We will closely follow geometrical approach and the conventions of \cite{Faraggi:2011ge}. In particular, we rely on App. B of \cite{Faraggi:2011ge}, where a summary of the geometric structure of embedded manifolds is given. See also \cite{Forini:2015mca} for a similar approach. In what follows, target-space indices are denoted by $m,n,\ldots$, world-sheet indices are $a,b,\ldots$, while the directions orthogonal to the string are represented by $i,j,\ldots$. All corresponding tangent space indices are underlined.

\subsection{Type IIB strings on $AdS_5\times S^5$}

In the bosonic sector, the string dynamics is dictated by the Nambu-Goto (NG) action
\begin{empheq}{align}
	S_\textrm{NG}&=\frac{1}{2\pi \alpha'}\int d^2\sigma\,\sqrt{g}\,,
\end{empheq}
where $g_{ab}$ is the induced metric on the world sheet and $g=\left|\det g_{ab}\right|$. Our first goal in this section is to consider perturbations $x^m\rightarrow x^m+\delta x^m$ around any given classical embedding and to find the quadratic action that governs them. To this purpose, let us choose convenient vielbeins for the $AdS_5\times S^5$ metric that are properly adapted to the study of fluctuations. Using the local $SO(9,1)$ symmetry, we can always pick a frame $E^{\underline{m}}=(E^{\underline{a}},E^{\underline{i}})$ such that the pullback of $E^{\underline{a}}$ onto the world-sheet forms a vielbein for the induced metric, while the pullback of $E^{\underline{i}}$ vanishes. Of course, these are nothing but the 1-forms dual to the tanget and normal vectors fields, respectively. The Lorentz symmetry is consequently broken to $SO(1,1)\times SO(8)$. Having made this choice, we may define the fields
\begin{empheq}{align}
\label{Eq:Fluctuations}
	\chi^{\underline{m}}&=E^{\underline{m}}_{\phantom{\underline{m}}m}\delta x^m\,,
\end{empheq}
and gauge fix the diffeomorphism invariance by freezing the tangent fluctuations, namely, by requiring
\begin{empheq}{align}
\label{Eq:VielGauge}
	\chi^{\underline{a}}&=0\,.
\end{empheq}
The physical degrees of freedom are then parameterized by the normal directions $\chi^{\underline{i}}$. This choice has the advantage that the gauge-fixing determinant is trivial \cite{Drukker:2000ep}. In this gauge, the variation of the induced metric is
\begin{empheq}{align}
	\delta g_{ab}&=-2H_{\underline{i}ab}\chi^{\underline{i}}+
	\nabla_a\chi^{\underline{i}}\nabla_b\chi^{\underline{j}}\delta_{\underline{ij}}
	+\left(H_{\underline{i}a}^{\phantom{\underline{i}a}c}H_{\underline{j}bc}
	-R_{m\underline{i}n\underline{j}}\partial_ax^m\partial_bx^n\right)\chi^{\underline{i}}\chi^{\underline{j}}\,,
\end{empheq}
where $H^{\underline{i}}_{\phantom{\underline{i}}ab}$ is the extrinsic curvature of the embedding and
\begin{empheq}{alignat=5}
	 \nabla_a\chi^{\underline{i}}&=\partial_a\chi^{\underline{i}}+\mathcal{A}^{\underline{ij}}_{\phantom{\underline{ij}}a}\chi_j
\end{empheq}
is the world-sheet covariant derivative, which includes the $SO(8)$ normal bundle connection $\mathcal{A}^{\underline{ij}}_{\phantom{\underline{ij}}a}$. These objects, as well as the world-sheet spin connection $w^{\underline{ab}}$, are related to the pullback of the target-space spin connection $\Omega^{\underline{mn}}$ by
\begin{empheq}{alignat=5}
	w^{\underline{ab}}&=P[\Omega^{\underline{ab}}]\,,
	&\qquad
	 H^{\underline{i}}_{\phantom{\underline{i}}ab}&=P[\Omega^{\underline{i}}_{\phantom{\underline{i}}\underline{a}}]_ae^{\underline{a}}_{\phantom{\underline{a}}b}\,,
	&\qquad
	\mathcal{A}^{\underline{ij}}&=P[\Omega^{\underline{ij}}]\,,
\end{empheq}
where $e^{\underline{a}}_{\phantom{\underline{a}}a}=P\left[E^{\underline{a}}\right]_a$ is the induced geometry vielbein. Using the well-known expansion of the square root of a determinant, a short calculation shows that, to quadratic order, the NG action becomes
\begin{empheq}{align}
\label{Eq:S2NG}
	S^{(2)}_\textrm{NG}&=\frac{\sqrt{\lambda}}{4\pi}\int d\tau d\sigma\,\sqrt{g}\left(g^{ab}\nabla_a\chi^{\underline{i}}\nabla_b\chi^{\underline{j}}\delta_{\underline{ij}}-\left(g^{ab}H_{\underline{i}a}^{\phantom{\underline{i}a}c}H_{\underline{j}bc}+\delta^{\underline{ab}}R_{\underline{aibj}}\right)\chi^{\underline{i}}\chi^{\underline{j}}\right)\,,
\end{empheq}
where we have used the equations of motion $g^{ab}H^{\underline{i}}_{\phantom{\underline{i}}ab}=0$ and written $g^{ab}R_{m\underline{i}n\underline{j}}\partial_ax^m\partial_bx^n=\delta^{\underline{ab}}R_{\underline{aibj}}$. We have traded the string tension for the 't Hooft coupling of the gauge theory, using $\lambda=1/\alpha'^2$. The continuation of this expression to Euclidean signature is straightforward.

Let us now discuss the fermionic degrees of freedom. In Lorentzian signature, the type IIB string involves a doublet of 10-dimensional positive chirality Majorana-Weyl spinors, $\theta_I$. At quadratic order, the Green-Schwarz (GS) action that controls their dynamics on $AdS_5\times S^5$ is given by \cite{Drukker:2000ep,Metsaev:1998it}
\begin{empheq}{alignat=5}\label{Eq:FermionsGS}
	S_\textrm{GS}&=\frac{i\sqrt{\lambda}}{2\pi}\int d\tau d\sigma\,\left(\sqrt{g}g^{ab}\delta^{IJ}-\epsilon^{ab}s^{IJ}\right)\overline{\theta}_I\Gamma_a\left(D_b\theta\right)_J\,,
\end{empheq}
where $s^{11}=-s^{22}=1$, $s^{12}=s^{21}=0$, the symbol $\epsilon^{ab}$ is a density with $\epsilon^{01}=1$ and $\Gamma_a=\Gamma_m\partial_ax^m$ is the pullback of the 10-dimensional Dirac matrices. Also, $D_a=\partial_ax^mD_m$ is the pullback of the spacetime covariant derivative appearing in the supersymmetry variation of the gravitino, which includes the contribution from the RR 5-form. Explicitly \cite{Martucci:2005rb}
\begin{empheq}{alignat=5}
	D^{IJ}_m&=\nabla_m\delta^{IJ}+\frac{1}{16\cdot5!}F^{nopqr}\Gamma_{nopqr}\Gamma_m\epsilon^{IJ}\,.
\end{empheq}

The above action can be simplified considerably. Indeed, given our choice of vielbein, we have
\begin{empheq}{align}
	 \partial_ax^mD_m^{IJ}&=\nabla_a\delta^{IJ}-\frac{1}{2}H^{\underline{i}\phantom{a}\underline{a}}_{\phantom{\underline{i}}a}\Gamma_{\underline{ai}}\delta^{IJ}+\frac{1}{16}\slashed{F}_5\Gamma_a\epsilon^{IJ}\,,
\end{empheq}
where the world-sheet covariant derivative $\nabla_a$ includes the normal bundle connection $\mathcal{A}^{\underline{ij}}_{\phantom{\underline{ij}}a}$, that is,
\begin{empheq}{align}
	 \nabla_a&=\partial_a+\frac{1}{4}w^{\underline{ab}}_{\phantom{\underline{ab}}a}\Gamma_{\underline{ab}}+\frac{1}{4}\mathcal{A}^{\underline{ij}}_{\phantom{\underline{ij}}a}\Gamma_{\underline{ij}}\,.
\end{empheq}
Using the relation $\epsilon^{ab}\Gamma_a=\sqrt{g}\,\Gamma_{\underline{01}}\Gamma^b$, it is easy to see that the terms proportional to the extrinsic curvature drop out from the action because of the equations of motion $H^{\underline{i}}_{\phantom{\underline{i}}ab}\Gamma^a\Gamma^b=H^{\underline{i}}_{\phantom{\underline{i}}ab}g^{ab}=0$. Then,
\begin{empheq}{alignat=5}
	S_\textrm{GS}&=\frac{i\sqrt{\lambda}}{2\pi}\int d\tau d\sigma\sqrt{g}\,\overline{\theta}_I\left(\delta^{IJ}-s^{IJ}\Gamma_{\underline{01}}\right)\left(\delta_J^{\phantom{J}K}\Gamma^a\nabla_a+\frac{1}{16}\epsilon_J^{\phantom{J}K}\Gamma^a\slashed{F}_5\Gamma_a\right)\theta_K\,.
\end{empheq}
Finally, notice that, in addition to diffeomorphism invariance and local Lorentz rotations, the GS action also enjoys the local $\kappa$-symmetry
\begin{empheq}{alignat=5}
	\delta\theta_I&=\frac{1}{2}\left(\delta_I^{\phantom{I}J}-s_I^{\phantom{I}J}\Gamma_{\underline{01}}\right)\kappa_J\,.
\end{empheq}
It is then possible to gauge fix to $\theta_1=\theta_2\equiv\theta$, as done in \cite{Drukker:2000ep}. This results in
\begin{empheq}{alignat=5}\label{eq: GS Lorentzian kappa-fixed}
	S_\textrm{GS}&=\frac{i\sqrt{\lambda}}{\pi}\int d\tau d\sigma\sqrt{g}\,\overline{\theta}\left(\Gamma^a\nabla_a-\frac{1}{16}\Gamma_{\underline{01}}\Gamma^a\slashed{F}_5\Gamma_a\right)\theta\,.
\end{empheq}

\subsection{Spectrum of excitations}

Let us now specialize the above results to the case of interest. All geometric ingredients needed to evaluate the actions have been collected in App.~\ref{App: geometric data}, while the dimensional reduction of the spinor $\theta$ is carried out in detail in App.~\ref{App: Dimensional reduction}. We will work exclusively in Euclidean signature.

For the bosonic fluctuations $\chi^{\underline{i}}$, we find that the quadratic action ruling them is given by (all fields are generically denoted by $\chi$)
\begin{empheq}{align}
	\label{eq: bosonic action 234}
	S^{2,3,4}&=\frac{\sqrt{\lambda}}{4\pi}\int d\tau d\sigma\,\sqrt{g}\left(g^{ab}\partial_a\chi\partial_b\chi+\frac{2}{\sqrt{g}}\chi^2\right)\,,
	\\
	\label{eq: bosonic action 56}
	S^{5,6}&=\frac{\sqrt{\lambda}}{2\pi}\int d\tau d\sigma\,\sqrt{g}\left(g^{ab}D_a\chi(D_b {\chi})^\dagger-\frac{2m^2}{\sqrt{g}}|\chi|^2\right)\,,
	\\
	\label{eq: bosonic action 789}
	S^{7,8,9}&=\frac{\sqrt{\lambda}}{4\pi}\int d\tau d\sigma\,\sqrt{g}\left(g^{ab}\partial_a\chi\partial_b\chi-\frac{2\sin^2\theta}{\sqrt{g}}\,\chi^2\right)\,.
\end{empheq}
In the second line, $\chi$ is a complex scalar field defined as $\chi=\frac{1}{\sqrt{2}}\left(\chi^{\underline{5}}+i\chi^{\underline{6}}\right)$, and the $\sigma$-dependent mass term reads
\begin{empheq}{alignat=5}
	m&=\frac{\sinh\rho\sin\theta\left(\cosh\rho-\cos\theta\right)}{A}&\,=\frac{1}{\cosh\left(2\sigma+\sigma_0\right)}\,,
\end{empheq}
where $A$ is the conformal factor in (\ref{A}). The covariant derivative also includes a $U(1)$ connection $\mathcal{A}$, namely,
\begin{empheq}{align}
	D_a\chi&=\partial_a\chi-i\mathcal{A}_a\chi\,,
\end{empheq}
with\footnote{This corresponds to the choice $\delta(\tau)=\tau$ in App.~\ref{App: geometric data}.}
\begin{empheq}{alignat=5}
	\mathcal{A}&=\left(\frac{\sinh^2\rho\cos\theta+\cosh\rho\sin^2\theta}{A}-1\right)d\tau&\,=\left(\tanh\left(2\sigma+\sigma_0\right)-1\right)d\tau\,.
\end{empheq}
Notice that $\mathcal{A}$ is regular at the center of the disk $\sigma\rightarrow\infty$ thanks to the $-1$ in the above expression. This is the reason why we have chosen this particular gauge. In what follows we will abuse notation and call $\mathcal{A}_{\tau}=\mathcal{A}$.

A few comments are in order. First, the $SO(3)\times SO(2)\times SO(3)$ invariance of the bosonic spectrum follows directly from the structure of equations \eqref{eq: bosonic action 234}, \eqref{eq: bosonic action 56} and \eqref{eq: bosonic action 789}. Second, we notice that, due to Weyl invariance, the action for the fluctuations $\chi^{\underline{2},\underline{3},\underline{4}}$ corresponds to a standard scalar field action in $AdS_2$ with mass term $m^2=2$ (in units of the $AdS$ radius). Third, the mass terms for $\chi^{\underline{5},\underline{6}}$ and $\chi^{\underline{7},\underline{8},\underline{9}}$ all vanish in the limit $\theta_0\rightarrow0$, and so does the gauge field, thus recovering the
$SL(2,\mathbb{R})\times SO(3)\times SO(5)\subset OSp(4^*|4)$ bosonic symmetry of the 1/2 BPS solution, which has been worked out explicitly in \cite{faraggi:2011bb}. After a unitary transformation, this spectrum is in agreement with the one found in \cite{Forini:2015bgo}.

Let us now move on to the fermionic fields. When applying the formalism above to the string dual to the 1/4 BPS Wilson loop, we are faced with a subtle issue. The classical world-sheet is Euclidean regardless of the signature of the target space. The GS action, however, is only defined for a Lorentzian metric. We will take a pragmatic approach and formally continue the fermionic action to a Euclidean world-sheet. Also, we shall compute all the relevant geometric quantities using a Euclidean $AdS_5\times S^5$ vielbein. The main drawback is that, since the Majorana condition on the spinors must be dropped, the action ceases to be real. Despite this fact, we find it convenient to proceed in this way in order to avoid further contrivances. The continuation of \eqref{eq: GS Lorentzian kappa-fixed} gives
\begin{empheq}{alignat=5}\label{eq: GS Euclidean kappa-fixed}
	S_\textrm{ferm}&=\frac{\sqrt{\lambda}}{\pi}\int d\tau d\sigma\sqrt{g}\,\overline{\theta}\left(\Gamma^a\nabla_a-\frac{i}{16}\Gamma_{\underline{01}}\Gamma^a\slashed{F}_5\Gamma_a\right)\theta\,,
\end{empheq}
where all world-sheet and target space quantities are intrinsically Euclidean, including the RR flux, which is now complex.

After dimensionally reducing the spinor $\theta$ according to the $SO(2)\times SU(2)\times U(1)\times SU(2)\subset SO(10)$ decomposition detailed in App.~\ref{App: Dimensional reduction}, we end up with eight 2-dimensional Dirac spinors $\psi^{\alpha'\alpha''}_{\alpha}$. The labels $(\alpha,\alpha',\alpha'')=(\pm,\pm,\pm)$ carry the $U(1)\times SU(2)\times SU(2)\subset SU(2|2)$ representations of the fields. Equation \eqref{eq: GS Euclidean kappa-fixed} then dictates that each of these fluctuations is governed by the action (all indices in $\psi^{\alpha'\alpha''}_{\alpha}$ are being hidden)
\begin{empheq}{alignat=5}\label{eq: fermionic action}
	S^{\alpha}_\textrm{ferm}&=\frac{\sqrt{\lambda}}{\pi}\int d\tau d\sigma\,\sqrt{g}\,\overline{\psi}\left(\gamma^a\nabla_a-\frac{1}{\sqrt{g}}\left(\gamma_{\underline{01}}\sinh^2\rho+i\alpha\sin^2\theta\right)\right)\psi\,,
\end{empheq}
where the covariant derivative is now
\begin{empheq}{alignat=5}
	 \nabla_a\psi&=\partial_a\psi+\frac{1}{4}w^{\underline{ab}}_{\phantom{\underline{ab}}a}\gamma_{\underline{ab}}\psi+i\frac{\alpha}{2}\mathcal{A}_a\,\psi\,.
\end{empheq}
Notice that the only label that matters in the above expressions is the $U(1)$ charge $\alpha=\pm$. The field content is therefore captured by four copies of each species of fermions. The invariance of the action under $U(1)\times SU(2)\times SU(2)\subset SU(2|2)$ is manifest since all (hidden) indices are properly contracted. In fact, as we shall see momentarily, the total action is invariant under the full supergroup $SU(2|2)$.

\subsection{Multiplet structure and supersymmetry}
\label{sec: multiplet}

Before dropping the labels $\alpha'$ and $\alpha''$ for the reminder of the paper, let us comment on how the full spectrum of fluctuations fits into supermultiplets of the supergroup $SU(2|2)$ preserved by the latitude background. A relevant reference on this matter is given by \cite{Beisert:2006qh}, see also \cite{BahaBalantekin:1980tup,BahaBalantekin:1980igq,BahaBalantekin:1981fmm}. 

It is useful to think in terms of the bosonic subgroup $SU(2)\times SU(2)$ of $SU(2|2)$. A generic long multiplet $[m,n]_q$, labeled by two natural numbers $m$ and $n$ and the $U(1)$ central charge $q$, decomposes as (see eq. (2.8) of \cite{Beisert:2006qh})
\begin{empheq}{align}\label{eq: multiplet mn}
	[m,n]_q&=\left\{
	\begin{array}{cccc}
		(m+0,n+0) & (m+0,n+0) & (m+0,n+0) & (m+0,n+0) \\
		(m+2,n+0) & (m+0,n+2) & (m-2,n+0) & (m+0,n-2) \\
		\hline
		(m+1,n+1) & (m+1,n+1) & (m-1,n+1) & (m-1,n+1) \\
		(m+1,n-1) & (m+1,n-1) & (m-1,n-1) & (m-1,n-1) \\
	\end{array}
	\right\}\,,
\end{empheq}
where $(p,q)$ specify the Dynkin labels of $SU(2)\times SU(2)$. Each of these labels is equal to twice the corresponding spin. The upper (lower) two lines represent bosonic (fermionic) components, all of which have the same $U(1)$ charge. The dimension of the representation is $16(m+1)(n+1)$. For small values of $m$ and $n$ one has a slightly different expression since some components in \eqref{eq: multiplet mn} are absent. In particular, for $m=n=0$, which is, as we will see below, the case that interests us, the multiplet reads
\begin{empheq}{align}\label{eq: multiplet 000}
	[0,0]_q&=\left\{
	\begin{array}{cc}
		(0,0) & (0,0) \\
		(2,0) & (0,2) \\
		\hline
		(1,1) & (1,1)
	\end{array}
	\right\}\,.
\end{empheq}
This representation has dimension $16=8+8$.

Looking at the bosonic spectrum and the way that the $SU(2)\times SU(2)\simeq SO(3)\times SO(3)$ symmetry is realized geometrically as a residual global symmetry of the local $SO(10)$ rotations, we see that the set of fields $\left\{\chi^{\underline{2}},\chi^{\underline{3}},\chi^{\underline{4}}\right\}$ transforms as a triplet under the first $SU(2)$ factor and as a singlet under the second factor, {\it i.e.} $(p,q)=(2,0)$. Similarly, $\left\{\chi^{\underline{7}},\chi^{\underline{8}},\chi^{\underline{9}}\right\}$ belong, all together, to $(p,q)=(0,2)$. Finally, $\chi^{\underline{5}}$ and $\chi^{\underline{6}}$ each have $(p,q)=(0,0)$. This is precisely the structure encoded above the solid line in \eqref{eq: multiplet 000}. As for the fermions $\psi^{\alpha'\alpha''}_{\alpha}$, the analysis in App. \ref{App: Dimensional reduction} shows that the labels $\alpha'$ and $\alpha''$ each correspond to a spin-$\frac{1}{2}$ representation of $SU(2)$, namely, $(p,q)=(1,1)$.

Now, to study the $U(1)$ charge assignments in the spectrum we must consider the Fourier expansions $\chi(\tau,\sigma)=e^{iE\tau}\chi_E(\sigma)$ and $\psi(\tau,\sigma)=e^{iE\tau}\psi_E(\sigma)$. Bosonic fields have integer $E$. For fermions, on the other hand, $E$ must be a half-integer in order to comply with the only allowed spin structure on a smooth manifold with a contractible cycle. This is mandatory in a gauge where all the fields are regular at the center of the disk $\sigma\rightarrow\infty$, which is indeed our case.\footnote{The fact that we will introduce a large $\sigma$ regulator, $R$, means that we are effectively removing the origin from the disk, which would allow for periodic fermions. However, this spin structure is unnatural considering that the $R\rightarrow\infty$ limit is eventually taken.}

By definition, any field $\phi$ of charge $q$ behaves like $\phi\rightarrow e^{iq\lambda}\phi$ under a $U(1)$ transformation with parameter $\lambda$. In this case, the symmetry is implemented by a shift $\tau\rightarrow\tau+\Delta\tau$, corresponding to an isometry of the world-sheet geometry, complemented by a rotation of the $5$-$6$ plane by an angle $\Delta\tau$. Any given Fourier mode will have a contribution to its $U(1)$ charge coming from the fact that $e^{iE\tau}\rightarrow e^{iE\Delta\tau}e^{iE\tau}$. Moreover, the scalars $\chi^{\underline{5}}$ and $\chi^{\underline{6}}$, as well as the fermions, are affected by the rotation in the $5$-$6$ plane via a phase proportional to the gauge field coupling appearing in the covariant derivative. Thus, we find the following set of charges:
\begin{empheq}{align}
	\begin{array}{|c|c|c|c|}
		\hline
		\textrm{Fields} & \multicolumn{2}{c|}{U(1)} & SU(2)\times SU(2)
		\\
		\hline
		\phantom{\bigg|} \chi_E^{\underline{2}},\chi_E^{\underline{3}},\chi_E^{\underline{4}}\phantom{\bigg|} & E & \multirow{3}{*}{$E\in\mathds{Z}$} & (2,0)
		\\
		\cline{1-2}\cline{4-4}
		\phantom{\bigg|}\chi_E^{\underline{5}}\pm i\chi_E^{\underline{6}}\phantom{\bigg|} & E\pm1 &  & (0,0)
		\\
		\cline{1-2}\cline{4-4}
		\phantom{\bigg|}\chi_E^{\underline{7}},\chi_E^{\underline{8}},\chi_E^{\underline{9}}\phantom{\bigg|} & E & & (0,2)
		\\
		\hline
		\phantom{\bigg|}{\psi^{\alpha'\alpha''}_{\pm}}_E\phantom{\bigg|} & E\mp\frac{1}{2} & E\in\mathds{Z}+\frac{1}{2} & (1,1)
		\\
		\hline
	\end{array}\,.
\end{empheq}
Notice that all the fields have integer charge. We can fit the Fourier components into multiplets as follows
\begin{empheq}{align}\label{eq: multiplet 00}
	[0,0]_{E\,\in\,\mathds{Z}}&=\left\{
	\begin{array}{cc}
		 \left\{\chi_E^{\underline{2}},\chi_E^{\underline{3}},\chi_E^{\underline{4}}\right\}\oplus\left(\chi_{E-1}^{\underline{5}}+i\chi_{E-1}^{\underline{6}}\right)\oplus\left(\chi_{E+1}^{\underline{5}}-i\chi_{E+1}^{\underline{6}}\right)\oplus\left\{\chi_E^{\underline{7}},\chi_E^{\underline{8}},\chi_E^{\underline{9}}\right\}
		 \\
		\hline
		\left\{{\psi^{\alpha'\alpha''}_+}_{E+\frac{1}{2}}\right\}\oplus\left\{{\psi^{\alpha'\alpha''}_-}_{E-\frac{1}{2}}\right\}
	\end{array}
	\right\}\,.
\end{empheq}
In terms of $SU(2|2)$ supermultiplets, the spectrum of excitations of the 1/4 BPS string dual to the latitude Wilson loop is then given by
\begin{empheq}{align}
	\bigoplus_{E\,\in\,\mathds{Z}}\;[0,0]_E\,.
\end{empheq}
This shows that the action for the quadratic fluctuations is invariant under the full $SU(2|2)$ supergroup.

\section{One-loop determinants}
\label{Sec:Determinants}

In this section we compute the ratio between the 1-loop determinants of the quadratic fluctuations around the string configurations corresponding to the 1/4 BPS latitude and  the 1/2 BPS circle. To this scope, we shall employ the Gelfand-Yaglom (GY) method \cite{Gelfand:1959nq}. See \cite{Kirsten:2007ev} for a  pedagogical review and \cite{Kruczenski:2008zk} for its application to the computation of the 1/2 BPS Wilson loop effective action. This method was also recently used in \cite{Forini:2015bgo} to compute the same ratio we are considering. One difference with  respect to that reference is that we will consider the first order Dirac-like fermionic operator as opposed to the second order one that results from squaring it. This will allow us to obtain analytic results and to avoid having to resort to numerics. Moreover, we consider the ratio between individual modes, rather than the ratio between the full 1/4 BPS and 1/2 BPS determinants. In order to regulate divergences we introduce two regulators for small and large $\sigma$ that we call $\epsilon_0$ and $R$ and that will be sent to 0 and $\infty$, respectively.

The path integral over the fluctuations yields the formal result
\begin{empheq}{align}
	e^{-\Gamma_{\textrm{effective}}^{1-\textrm{loop}}}&=\frac{\left(\textrm{Det}\,\mathcal{O}^+\right)^{\frac{4}{2}}\,\left(\textrm{Det}\,\mathcal{O}^-\right)^{\frac{4}{2}}}{\left(\textrm{Det}\,\mathcal{O}^{2,3,4}\right)^{\frac{3}{2}}\,\left(\textrm{Det}\,\mathcal{O}^{5,6}\right)^{\frac{2}{2}}\,\left(\textrm{Det}\,\mathcal{O}^{7,8,9}\right)^{\frac{3}{2}}}\,,
\end{empheq}
where the differential operators follow from integration by parts in \eqref{eq: bosonic action 234}, \eqref{eq: bosonic action 56}, \eqref{eq: bosonic action 789}, and \eqref{eq: fermionic action}. To account for the Majorana nature of the type IIB spinors in Lorentzian signature, we have taken the square root of the fermionic operators. The fact that we have combined the fluctuations $\chi^{\underline{5}}$ and $\chi^{\underline{6}}$ into a single complex field has also been taken into consideration.

Notice that due to Weyl invariance the bosonic operators can be naturally defined with respect to the flat metric $\eta_{ab}$, which corresponds to a rescaling of the induced geometry operators by $\sqrt{g}$. Such a transformation is inconsequential at the level of the path integral as long as it is accompanied by the concomitant rescaling of the fermionic operators by $g^{1/4}$, this in order to cancel the conformal anomaly \cite{Drukker:2000ep}. In what follows we will always work with the rescaled version of the operators.

We shall proceed by making a Fourier expansion of the fields whereby $\partial_{\tau}\rightarrow iE$. Then, the determinant of any given two-dimensional operator, $\mathcal{O}$, can be computed as
\begin{empheq}{align}
	\ln\left(\textrm{Det}\,\mathcal{O}\right)&=\sum_E\ln\left(\textrm{Det}\,\mathcal{O}_E\right)\,,
\end{empheq}
where $\mathcal{O}_E$ is the corresponding one-dimensional operator acting on a specific Fourier mode. For the case at hand, the relevant one-dimensional differential operators are
\begin{empheq}{align}
	\label{eq: bosonic operators 234}
	\mathcal{O}_E^{2,3,4}&=-\partial_{\sigma}^2+E^2+2\sinh^2\rho\,,
	\\
	\label{eq: bosonic operators 56}
	\mathcal{O}_E^{5,6}&=-\partial_{\sigma}^2+\left(E-\mathcal{A}\right)^2-2m^2\,,
	\\
	\label{eq: bosonic operators 789}
	\mathcal{O}_E^{7,8,9}&=-\partial_{\sigma}^2+E^2-2\sin^2\theta
\end{empheq}
for the bosonic modes, and
\begin{empheq}{align}\label{eq: fermionic operators}
	\mathcal{O}_E^{\alpha}&=\gamma_{\underline{1}}\left(\partial_{\sigma}+\frac{1}{2}w\right)+i\gamma_{\underline{0}}\left(E+\frac{\alpha}{2}\mathcal{A}\right)+\frac{1}{\sqrt{A}}\left(\sinh^2\rho\gamma_{\underline{01}}+i\alpha\sin^2\theta\right)
\end{empheq}
for the fermions. Notice that $\gamma_{\underline{0}}\mathcal{O}^{\alpha}_E\gamma_{\underline{0}}=-\mathcal{O}^{-\alpha}_{-E}$, so the determinants in the two charged sectors, with appropriate boundary conditions, should coincide up to a phase. We will confirm this expectation below.

The above operators generically depend on the value of $\sigma_0$ that characterizes the classical string solution. We will define the ratios
\begin{empheq}{align}
	\label{eq: bosonic ratios 234}
	\Omega_E^{2,3,4}(\sigma_0)&=\ln\left[\frac{\textrm{Det}\,\mathcal{O}_E^{2,3,4}(\sigma_0)}{\textrm{Det}\,\mathcal{O}_E^{2,3,4}(\infty)}\right]\,,
	\\
	\label{eq: bosonic ratios 56}
	\Omega_E^{5,6}(\sigma_0)&=\ln\left[\frac{\textrm{Det}\,\mathcal{O}_E^{5,6}(\sigma_0)}{\textrm{Det}\,\mathcal{O}_E^{5,6}(\infty)}\right]\,,
	\\
	\label{eq: bosonic ratios 789}
	\Omega_E^{7,8,9}(\sigma_0)&=\ln\left[\frac{\textrm{Det}\,\mathcal{O}_E^{7,8,9}(\sigma_0)}{\textrm{Det}\,\mathcal{O}_E^{7,8,9}(\infty)}\right]\,,
\end{empheq}
and
\begin{empheq}{align}
	\Omega_E^{\alpha}(\sigma_0)&=\ln\left[\frac{\textrm{Det}\,\mathcal{O}_E^{\alpha}(\sigma_0)}{\textrm{Det}\,\mathcal{O}_E^{\alpha}(\infty)}\right]
\end{empheq}
between the determinants of the 1/4 BPS and 1/2 BPS operators. Each ratio is to be computed using the GY method. We emphasize that we are defining the ratio of determinants between the same set of modes of two different string configurations (the 1/4 BPS and the 1/2 BPS strings) and not the ratio between bosonic and fermionic modes within each individual solution. The advantages of doing this are manifold. First, given that the world-sheets have the same topology, we expect the divergences coming from the small $\sigma$ regulator, $\epsilon_0$, to cancel within each ratio. That is, each $\Omega(\sigma_0)$ should be finite as $\epsilon_0\rightarrow0$. Second, this allows us to work directly with the first order fermionic operators without the need for squaring them. This simplifies the computations considerably and allows for an analytic result.

The expression for the difference of the 1-loop effective actions between the 1/4 BPS and 1/2 BPS strings is then given by
\bea
	\Delta\Gamma_{\textrm{effective}}^{1-\textrm{loop}}(\sigma_0)&=&\frac{1}{2}\sum_{E\in\mathds{Z}}
	\left(3\Omega_E^{2,3,4}(\sigma_0)+2\Omega_E^{5,6}(\sigma_0)+3\Omega_E^{7,8,9}(\sigma_0)\right)
	-\frac{4}{2}\sum_{E\in\mathds{Z}+\frac{1}{2}}\left(\Omega_E^+(\sigma_0)+\Omega_E^-(\sigma_0)\right)\,.\cr &&
	\label{final-dif}
\eea
As mentioned before, a factor of $\frac{1}{2}$ for the fermionic modes is being introduced by hand in order to account for the Majorana condition which was lost in the Euclidean continuation of the GS action. We will later describe the specific procedure we followed for regulating and performing these sums. 

\subsection{The Gelfand-Yaglom method}
\label{sec:GY}

Here we briefly review the GY method applied to our particular string configurations, see also \cite{Forini:2015bgo}. Consider two general operators of the form
\begin{empheq}{align}
	\mathcal{O}&=P_0(\sigma)\frac{d^n}{d\sigma^n}+P_1(\sigma)\frac{d^{n-1}}{d\sigma^{n-1}}+\sum_{k=2}^nP_k(\sigma)\frac{d^{n-k}}{d\sigma^{n-k}}\,,
	\\
	\hat{\mathcal{O}}&=P_0(\sigma)\frac{d^n}{d\sigma^n}+\hat{P}_1(\sigma)\frac{d^{n-1}}{d\sigma^{n-1}}+\sum_{k=2}^n\hat{P}_k(\sigma)\frac{d^{n-k}}{d\sigma^{n-k}}\,,
\end{empheq}
where $P_k(\sigma)$ are $r\times r$ matrices. These operators act on $r$-tuplet functions $\chi_s(\sigma)$, $s=1,\,\ldots,\,r$, defined on the interval $[\epsilon_0,R]$. We want to compute the determinants subject to the boundary conditions
\begin{empheq}{align}
	M\left(
	\begin{array}{c}\chi(\epsilon_0)
		\\\frac{d}{d\sigma}\chi(\epsilon_0)
		\\\vdots
		\\\frac{d^{n-1}}{d\sigma^{n-1}}\chi(\epsilon_0)
	\end{array}
	\right)
	+
	N\left(
	\begin{array}{c}
		\chi(R)
		\\\frac{d}{d\sigma}\chi(R)
		\\\vdots\\
		\frac{d^{n-1}}{d\sigma^{n-1}}\chi(R)
	\end{array}
	\right)
	&=
	\left(
	\begin{array}{c}
		0\\0\\
		\vdots
		\\0
	\end{array}
	\right)\,,
\end{empheq}
where $M$ and $N$ are two constant $nr\times nr$ matrices. The GY method does not yield each determinant individually, but rather provides a concise formula for their ratio \cite{raey}
\begin{empheq}{align}\label{eq: GY formula}
	\frac{\textrm{Det}\,\mathcal{O}}{\textrm{Det}\,\hat{\mathcal{O}}}&=\frac{e^{\int_{\epsilon_0}^Rd\sigma\,\textrm{tr}\left[\mathcal{R}(\sigma)P_1(\sigma)P_0^{-1}(\sigma)\right]}}{e^{\int_{\epsilon_0}^Rd\sigma\,\textrm{tr}\left[\mathcal{R}(\sigma)\hat{P}_1(\sigma)P_0^{-1}(\sigma)\right]}}\frac{\textrm{det}\left(M+NY_{\mathcal{O}}(R)\right)}{\textrm{det}\left(M+NY_{\hat{\mathcal{O}}}(R)\right)}\,.
\end{empheq}
Here,
\begin{empheq}{align}
	Y(\sigma)&=\left(
	\begin{array}{cccc}
		\chi^{(1)}(\sigma) & \chi^{(2)}(\sigma) & \cdots & \chi^{(n)}(\sigma)
		\\
		\frac{d}{d\sigma}\chi^{(1)}(\sigma) & \frac{d}{d\sigma}\chi^{(2)}(\sigma) & \cdots & \frac{d}{d\sigma}\chi^{(n)}(\sigma)
		\\
		\vdots & \vdots & \ddots & \vdots
		\\
		\frac{d^(n-1)}{d\sigma^{n-1}}\chi^{(1)}(\sigma) & \frac{d^{n-1}}{d\sigma^{n-1}}\chi^{(2)}(\sigma) & \cdots & \frac{d^{n-1}}{d\sigma^{n-1}}\chi^{(n)}(\sigma)
	\end{array}
	\right)\,,
\end{empheq}
is the fundamental matrix which collects the $n$ linearly independent solutions to the equation $\mathcal{O}\chi^{(i)}(\sigma)=0$, $i=1,\ldots,n$,  with boundary conditions $Y(\epsilon_0)=\mathds{1}_{nr\times nr}$, and $\mathcal{R}$ is a projector that selects half of the eigenvalues of $P_0$. For operators of even order, $\mathcal{R}_{\pm}=\pm\frac{1}{2}\mathds{1}$, but for odd order the definition is more complicated. Fortunately, in this paper we will only encounter first order examples where $P_0^2=\mathds{1}$. Then, $\mathcal{R}_{\pm}=\frac{1}{2}\left(1\pm P_0\right)$. The choice of sign determines which half of the eigenvalues is selected and does not affect the final result.

It is important to mention that the condition $P_0=\hat{P}_0$ is crucial for the validity of the GY method. In this sense, the rescaling of the bosonic and fermionic operators  discussed previously turns out to be essential in the application of the technique to the comparison of the 1/4 BPS and 1/2 BPS string effective actions, which have different conformal factors. The functions $P_0(\sigma)$ would otherwise differ in the two cases, rendering the method inapplicable.

In the case of second order scalar operators with $P_1=0$ and with Dirichlet-Dirichlet (D-D) or Dirichlet-Neumann (D-N) boundary conditions, the GY formula \eqref{eq: GY formula} yields 
\begin{empheq}{align}
	\frac{\textrm{Det}\,\mathcal{O}}{\textrm{Det}\,\hat{\mathcal{O}}}&=\lim_{R\rightarrow\infty} \left\{
	\begin{array}{cc}
		{\displaystyle \frac{\chi(R)}{\hat{\chi}(R)}}\,,\qquad & \textrm{D-D}
		\\\\
		{\displaystyle \frac{\chi'(R)}{\hat{\chi}'(R)}}\,,\qquad  & \textrm{D-N}
	\end{array}
	\right.\,,
\end{empheq}
where $\chi(\sigma)$ is the unique solution to $\mathcal{O}\chi(\sigma)=0$ satisfying
\begin{empheq}{alignat=5}\label{eq: GY scalar conditions}
	\chi(\epsilon_0)&=0\,,
	&\qquad
	\chi'(\epsilon_0)&=1\,,
\end{empheq}
and similarly for $\hat{\chi}$. These expressions will be used for the bosonic modes. We will find that in all cases the function $\chi(\sigma)$ can be written as
\begin{empheq}{align}\label{eq: GY scalar}
	\chi(\sigma)&=\chi_1(\sigma)\chi_2(\epsilon_0)-\chi_1(\epsilon_0)\chi_2(\sigma)\,,
\end{empheq}
where $\chi_1(\sigma)$ and $\chi_2(\sigma)$ are the properly normalized, linearly independent solutions to the equations of motion. The fermionic case will be discussed in due course.

\subsection{Bosonic determinants}

The implementation of the GY method for functional determinants requires solving the equations of motion for the string fluctuations. We will now proceed to do so, starting with the bosonic operators \eqref{eq: bosonic operators 234}, \eqref{eq: bosonic operators 56}, and \eqref{eq: bosonic operators 789} in order to compute the corresponding ratios in \eqref{eq: bosonic ratios 234}, \eqref{eq: bosonic ratios 56}, and \eqref{eq: bosonic ratios 789}. We assume D-D boundary conditions in the interval $[\epsilon_0,R]$, except for those modes $E$ that exhibit a special behavior at $R\rightarrow\infty$, for which D-N boundary conditions are to be imposed.

\subsubsection{Determinant for the $\chi^{2,3,4}$ modes}

For this group of fields we have the following equation (here and in the following we denote by $\chi$ the field of interest, suppressing the field label)
\begin{empheq}{alignat=5}
	-\partial_{\sigma}^2 \chi+E^2 \chi+\frac{2}{\sinh^2\sigma} \chi&=0\,,
\label{sep}
\end{empheq}
which is solved by
\begin{empheq}{alignat=5}
	\chi_1(\sigma)&=E\sinh(E\sigma)-\coth\sigma\cosh(E\sigma)\,,
	&\qquad
	\chi_2(\sigma)&=\frac{E\cosh(E\sigma)-\coth\sigma\sinh(E\sigma)}{E(E^2-1)}\,.
\end{empheq}
The normalization is chosen so that both functions survive the $E\rightarrow0$ and $E\rightarrow\pm1$ limits as linearly independent solutions. Furthermore, defining $\chi(\sigma)$ as in \eqref{eq: GY scalar}, one can verify that the conditions in \eqref{eq: GY scalar conditions} are indeed satisfied. Taking the $R\rightarrow\infty$ expansion, we find
\begin{empheq}{align}
	\chi(R)&\underset{R\rightarrow\infty}{\longrightarrow}\left\{
\begin{array}{ll}
	{\displaystyle\frac{e^{|E|\left(R-\epsilon_0\right)}}{2|E|\left(|E|+1\right)}\left(|E|+\coth\epsilon_0\right)}\,, &\qquad  E\neq0
	\\\\
	{\displaystyle R\coth\epsilon_0}\,, & \qquad E=0
\end{array}
\right.\,,
\end{empheq}
where we have kept all expressions exact in $\epsilon_0$. These expressions do not depend on the parameter $\sigma_0$. As a consequence, the ratio with the 1/2 BPS limit $\sigma_0\rightarrow\infty$ is trivial and gives
\begin{empheq}{align}
	\Omega_E^{2,3,4}(\sigma_0)&=0\,.
	\label{O234}
\end{empheq}

\subsubsection{Determinant for the $\chi^{5,6}$ modes}

These fluctuations satisfy the equation
\begin{empheq}{alignat=5}
	-\partial_{\sigma}^2\chi+\left(\left(E-\mathcal{A}\right)^2-2m^2\right)\chi&=0\,,
\label{sep2}
\end{empheq}
which can be recast, using that $2m^2=\partial_{\sigma}\mathcal{A}$, as
\begin{empheq}{align}
	\left[-\left(\partial_{\sigma}\mathcal{W}\right)^2+\partial^2_{\sigma}\mathcal{W}+\partial_{\sigma}^2\right]\chi&=0\,.
\end{empheq}
The prepotential is given by
\begin{empheq}{align}
	\partial_{\sigma}\mathcal{W}&=-E+\mathcal{A}\,=-E-1+\tanh\left(2\sigma+\sigma_0\right)\,,
	\cr
	\mathcal{W}&=-\frac{1}{2}\left(E+1\right)\left(2\sigma+\sigma_0\right)+\frac{1}{2}\ln\cosh\left(2\sigma+\sigma_0\right)\,.
\end{empheq}
We find that the two linearly independent solutions are
\begin{empheq}{align}
	\chi_1(\sigma)&=\frac{e^{\left(E+1\right)\left(\sigma+\sigma_0/2\right)}}{\sqrt{\cosh(2\sigma+\sigma_0)}}\,,
	\cr
	\chi_2(\sigma)&=\frac{e^{-\left(E+1\right)\left(\sigma+\sigma_0/2\right)}}{\sqrt{\cosh(2\sigma+\sigma_0)}}\left(
\frac{\left(E+1\right)\cosh(2\sigma+\sigma_0)+\sinh(2\sigma+\sigma_0)}{2E(E+2)}\right)-\frac{(E+1)\,G_1(\sigma)}{2E(E+2)}\,.
\end{empheq}
Again, $\chi_1(\sigma)$ and $\chi_2(\sigma)$ are finite and independent when $E\rightarrow0$ and $E\rightarrow-2$. The combination \eqref{eq: GY scalar} satisfies \eqref{eq: GY scalar conditions}. The relevant asymptotic expansions read
\begin{empheq}{align}
	\chi(R)&\underset{R\rightarrow\infty}{\longrightarrow}\left\{
\begin{array}{ll}
	{\displaystyle\frac{e^{E\left(R-\epsilon_0\right)}}{2E\left(E+2\right)}\sqrt{\frac{2}{1+\tanh\left(2\epsilon_0+\sigma_0\right)}}\left(E+1+\tanh\left(2\epsilon_0+\sigma_0\right)\right)}\,, \quad& E>0
	\\\\
	{\displaystyle R\sqrt{\frac{1+\tanh\left(2\epsilon_0+\sigma_0\right)}{2}}}\,, \quad & E=0
	\\\\
	{\displaystyle-\frac{e^{-E\left(R-\epsilon_0\right)}}{2E}\sqrt{\frac{1+\tanh\left(2\epsilon_0+\sigma_0\right)}{2}}}\,, \quad & E<0
\end{array}
\right.\,,
\end{empheq}
leading to
\begin{empheq}{align}
	\Omega_E^{5,6}(\sigma_0)&=\left\{
\begin{array}{ll}
	{\displaystyle-\ln\sqrt{\frac{1+\tanh\left(2\epsilon_0+\sigma_0\right)}{2}}+\ln\frac{E+1+\tanh\left(2\epsilon_0+\sigma_0\right)}{E+2}}\,,\qquad & E>0
	\\\\
	{\displaystyle \ln\sqrt{\frac{1+\tanh\left(2\epsilon_0+\sigma_0\right)}{2}}}\,,\qquad & E\leq0
	\label{Omega56}
\end{array}
\right.\,,
\end{empheq}
after one takes the ratio with the 1/2 BPS limit. We have checked that the special mode $E=0$ satisfies
\begin{empheq}{align}
	\lim_{R\rightarrow\infty}\frac{\chi(R)}{\underset{\sigma_0\rightarrow\infty}{\lim}\chi(R)}&=\lim_{R\rightarrow\infty}\frac{\chi'(R)}{\underset{\sigma_0\rightarrow\infty}{\lim}\chi'(R)}\,,
\end{empheq}
so the answer is unaffected by the choice of D-D or D-N boundary conditions.

\subsubsection{Determinant for the $\chi^{7,8,9}$ modes}

Finally, the field equation for the remaining fluctuations reads
\begin{empheq}{alignat=5}
	E^2\chi-\partial_{\sigma}^2\chi-\frac{2}{\cosh^2(\sigma+\sigma_0)}\chi&=0\,.
\label{sep3}
\end{empheq}
This has also simple solutions
\begin{empheq}{alignat=5}
	\chi_1(\sigma)&=\frac{E\sinh\left(E(\sigma+\sigma_0)\right)-\tanh(\sigma+\sigma_0)
\cosh\left(E(\sigma+\sigma_0)\right)}{E^2-1}\,,
	\cr
	 \chi_2(\sigma)&=\frac{E\cosh\left(E(\sigma+\sigma_0)\right)-\tanh(\sigma+\sigma_0)\sinh\left(E(\sigma+\sigma_0)\right)}{E}\,.
\end{empheq}
As before, the $E\rightarrow0$ and $E\rightarrow\pm1$ limits are well-defined leading to linearly independent functions, and the solution \eqref{eq: GY scalar} complies with the requirements \eqref{eq: GY scalar conditions}. One can verify that
\begin{empheq}{align}
	\chi(R)&\underset{R\rightarrow\infty}{\longrightarrow}\left\{
\begin{array}{ll}
	{\displaystyle\frac{e^{|E|\left(R-\epsilon_0\right)}}{2|E|(|E|+1)}\left(|E|+\tanh\left(\epsilon_0+\sigma_0\right)\right)}\,,\quad  & E\neq0, 
	\\\\
	{\displaystyle R\tanh\left(\epsilon_0+\sigma_0\right)}\,,\qquad  & E=0.
\end{array}
\right.\,.
\end{empheq}
Taking the ratio with the 1/2 BPS limit, one finds
\begin{empheq}{align}
	\Omega_E^{7,8,9}(\sigma_0)&=\ln\frac{|E|+\tanh\left(\epsilon_0+\sigma_0\right)}{|E|+1}\,.
	\label{Omega789}
\end{empheq}
As for the previous set of fluctuations, the special mode $E=0$ yields the same result for D-D or D-N boundary conditions.

\subsection{Fermionic determinants}

We now move on to study the fermionic degrees of freedom, whose equation of motion reads
\begin{empheq}{align}
	\left(\gamma_{\underline{1}}\left(\partial_{\sigma}+\frac{1}{2}w\right)+i\gamma_{\underline{0}}\left(E+\frac{\alpha}{2}\mathcal{A}\right)+\frac{1}{\sqrt{A}}\left(\sinh^2\rho\gamma_{\underline{01}}+i\alpha\sin^2\theta\right)\right)\psi&=0\,.
\end{empheq}
In order to simplify it to a point where we can solve it explicitly, we introduce the projectors
\begin{empheq}{align}
	P_{\pm}&=\frac{1}{2}\left(1\pm i\alpha\gamma_{\underline{01}}\right)\,,
\end{empheq}
and decompose
\begin{empheq}{alignat=5}
	\psi&=\psi_++\psi_-\,,
	&\qquad
	\psi_{\pm}&=P_{\pm}\psi\,.
\end{empheq}
Notice that these projections depend on the $U(1)$ charge $\alpha=\pm$, which we are omitting from the spinor $\psi$ in order to avoid confusion with the new $\pm$ labels in the equation above. The equation of motion in terms of these components splits as follows:
\begin{empheq}{align}
	\left(i\gamma_{\underline{0}}E+\gamma_{\underline{1}}D^{\pm}_{\sigma}\right)\psi_{\mp}+\frac{\gamma_{\underline{01}}}{\sqrt{A}}\left(\sinh^2\rho\mp\sin^2\theta\right)\psi_{\pm}&=0\,,
\end{empheq}
where $D^{\pm}_{\sigma}=\partial_{\sigma}+\frac{1}{2}w\pm\frac{1}{2}\mathcal{A}$. Solving for $\psi_-$, replacing it in the remaining equation, and using (\ref{wminusA}) we find
\begin{empheq}{align}
	\left(-\left(\partial_{\sigma}\mathcal{W}\right)^2+\partial^2_{\sigma}\mathcal{W}+\partial_{\sigma}^2\right)\psi_+&=0\,,
\end{empheq}
where the prepotential is
\bea
	\partial_{\sigma}\mathcal{W}&=&\alpha E-\frac{1}{2}\left(w-\mathcal{A}\right)
	=\alpha E-\frac{1}{2}+\frac{\cosh\left(2\sigma+\sigma_0\right)}{\sinh\left(2\sigma+\sigma_0\right)-\sinh\sigma_0}\,,
	\cr
	\mathcal{W}&=&\left(\alpha E-\frac{1}{2}\right)\left(\sigma+\frac{\sigma_0}{2}\right)+\frac{1}{2}\ln\left(\sinh\left(2\sigma+\sigma_0\right)-\sinh\sigma_0\right)\,.
\eea
This equation can be easily integrated, leading to the solution
\begin{empheq}{alignat=5}
	\psi_+(\sigma)&=I_1(\sigma)C_1+I_2(\sigma)C_2\,,
	&\qquad
	\psi_-(\sigma)&=\frac{\gamma_{\underline{0}}}{\sqrt{A}}\left(2I_1'(\sigma)C_1+\frac{\left(I_1(\sigma)I_2(\sigma)\right)'}{I_1(\sigma)}C_2\right)\,,
\end{empheq}
where $C_1$ and $C_2$ are two spinorial integration constants satisfying $P_+C_i=C_i$, and
\bea
	I_1(\sigma)&=&\frac{e^{-\left(\alpha E-\frac{1}{2}\right)\left(\sigma+\frac{\sigma_0}{2}\right)}}{\sqrt{\sinh(2\sigma+\sigma_0)-\sinh\sigma_0}}\,,
	\cr
	I_2(\sigma)&=&\frac{e^{\left(\alpha E-\frac{1}{2}\right)\left(\sigma+\frac{\sigma_0}{2}\right)}}{2\sqrt{\sinh(2\sigma+\sigma_0)-\sinh\sigma_0}}\left(\frac{\sinh\sigma_0}{\alpha E-\frac{1}{2}}+\frac{\cosh(2\sigma+\sigma_0)-\left(\alpha E-\frac{1}{2}\right)\sinh(2\sigma+\sigma_0)}{\left(\alpha E-\frac{1}{2}\right)^2-1}\right)
	\cr
	&&-\frac{1}{2}\left(\frac{\sinh\sigma_0}{\alpha E-\frac{1}{2}}+\frac{1}{\left(\alpha E-\frac{1}{2}\right)^2-1}\right)I_1(\sigma)\,.
\eea
These linear combinations survive the $\alpha E\rightarrow\pm\frac{1}{2}$ and $\alpha E\rightarrow\frac{3}{2}$ limits as independent functions. Notice that the interchange $\alpha \to -\alpha $ is equivalent to $E\to -E$. Also, the normalization has been chosen such that
\begin{empheq}{align}
	I_1'(\sigma)I_2(\sigma)-I_1(\sigma)I_2'(\sigma)&=1\,.
\end{empheq}

Let us now construct the fundamental matrix, $Y^{\alpha}(\sigma)$, for the fermionic operator. From now on, we will work in a basis where $\gamma_{\underline{0}}=\sigma_2,\,\gamma_{\underline{1}}=\sigma_1$ and $i\gamma_{\underline{01}}=\sigma_3$. Recalling the definition of the projectors $P_{\pm}$, this means that
\begin{empheq}{alignat=5}
	\psi&=\left(
	\begin{array}{c}
		\psi_+ \\
		\psi_-
	\end{array}
	\right)
	&\qquad
	\textrm{for $\alpha=1$}\,,
	&\qquad\qquad\qquad &
	\psi&=\left(
	\begin{array}{c}
		\psi_- \\
		\psi_+
	\end{array}
	\right)
	&\qquad
	\textrm{for $\alpha=-1$}\,.
\end{empheq}
We are slightly abusing notation here, since $\psi_{\pm}$ were defined as two-component spinors in the previous formulas. Now they represent specific components.

Starting with the case $\alpha=1$, we are instructed to find a $2\times2$ matrix, $Y^+(\sigma)$, of linearly independent solutions satisfying $Y^+(\epsilon_0)=\mathds{1}_{2\times2}$. One can check, using the above relations for $I_1$ and $I_2$, that the matrix
\begin{empheq}{align}
	Y^+(\sigma)&=\left(
	\begin{array}{cc}
		\psi_+^1(\sigma) & \psi_+^2(\sigma) \\
		\psi_-^1(\sigma) & \psi_-^2(\sigma)
	\end{array}
	\right)\bigg|_{\alpha=1}\,,
\end{empheq}
where
\bea
	\psi_+^1(\sigma)&=&\frac{I_1(\sigma)-2I_1'(\epsilon_0)I(\sigma)}{I_1(\epsilon_0)}\,,
	\qquad
	\psi_-^1(\sigma)=2i\frac{\left(I_1(\sigma)-2I_1'(\epsilon_0)I(\sigma)\right)I_1'(\sigma)-I_1(\epsilon_0)I_1'(\epsilon_0)}{\sqrt{A(\sigma)}I_1(\sigma)I_1(\epsilon_0)}\,,
	\cr
	\psi_+^2(\sigma)&=&-i\sqrt{A(\epsilon_0)}I(\sigma)\,,\qquad
	\psi_-^2(\sigma)=\sqrt{\frac{A(\epsilon_0)}{A(\sigma)}}\frac{2I_1'(\sigma)I(\sigma)+I_1(\epsilon_0)}{I_1(\sigma)}\,,
\eea
does the job. The function $I(\sigma)$ is given by
\begin{empheq}{align}
	I(\sigma)&=I_1(\sigma)I_2(\epsilon_0)-I_1(\epsilon_0)I_2(\sigma)\,,
\end{empheq}
and satisfies
\begin{empheq}{alignat=5}
	I(\epsilon_0)&=0\,,
	&\qquad
	I'(\epsilon_0)&=1\,.
\end{empheq}
For the other charged sector, namely $\alpha=-1$, the fundamental matrix is
\begin{empheq}{alignat=5}
	Y^-(\sigma)&=\gamma_{\underline{0}}Y^+(\sigma)\gamma_{\underline{0}}\bigg|_{E\rightarrow-E}
	&
	&=\left(
	\begin{array}{cc}
		\psi_-^2(\sigma) & -\psi_-^1(\sigma) \\
		-\psi_+^2(\sigma) & \psi_+^1(\sigma)
	\end{array}
	\right)\bigg|_{\alpha=-1}\,.
\end{empheq}
Notice that $\textrm{det}\,Y^{\alpha}(\sigma)=\sqrt{A(\epsilon_0)/A(\sigma)}$ is independent of $E$.

Next, we must compute the determinant of $M^{\alpha}+N^{\alpha}Y^{\alpha}(R)$, where the matrices $M^{\alpha}$ and $N^{\alpha}$ specify the boundary conditions for the fluctuations $\psi(\sigma)$ via $M^{\alpha}\psi(\epsilon_0)+N^{\alpha}\psi(R)=0$. 
We choose
\begin{empheq}{alignat=5}\label{eq: fermionic bc}
	M^{\alpha}&=\frac{1}{2}\left(A(\epsilon_0)\right)^{-\frac{1}{4}}\left(
	\begin{array}{cc}
		1+\alpha & 0 \\
		0 & 1-\alpha
	\end{array}
	\right)\,,
	&\qquad
	N^{\alpha}&=\frac{1}{2}\left(A(R)\right)^{-\frac{1}{4}}\left(
	\begin{array}{cc}
		0 & -1+\alpha \\
		1+\alpha & 0
	\end{array}
	\right)\,,
\end{empheq}
which implies $\psi_+(\epsilon_0)=\psi_+(R)=0$, whereas the other component $\psi_-$ remains unconstrained. Notice that $M^{\alpha}=\gamma_{\underline{0}}M^{-\alpha}\gamma_{\underline{0}}$ and $N^{\alpha}=\gamma_{\underline{0}}N^{-\alpha}\gamma_{\underline{0}}$. 
The prefactors in $M^\alpha$ and $N^\alpha$ can be justified by noticing that
\begin{empheq}{align}
	\mathcal{O}_E^{\alpha}&=e^{-\frac{1}{4}\ln A}\left[\gamma_{\underline{1}}\partial_{\sigma}+i\gamma_{\underline{0}}\left(E+\frac{\alpha}{2}\mathcal{A}\right)+\frac{1}{\sqrt{A}}\left(\sinh^2\rho\gamma_{\underline{01}}+i\alpha\sin^2\theta\right)\right]e^{\frac{1}{4}\ln A}\,,
\end{empheq}
so it is natural to impose the boundary conditions on $\tilde{\psi}(\sigma)=e^{\frac{1}{4}\ln A}\psi(\sigma)$ rather than on $\psi(\sigma)$ directly. Indeed, recalling that we have performed a conformal transformation so as to work with a flat metric, the fermionic fields respond precisely by acquiring the above prefactor \cite{Sakai:1984vm} and making the spin connection disappear from the operator. This will change the asymptotic behavior of the expressions involved in the GY formula. As will be commented on below, this rescaling of the boundary conditions is responsible for the cancellation of a linear $\Lambda$ divergence (but not of the $\ln\Lambda$ divergence, which cancels with or without the prefactors) that would otherwise appear when regulating the sum over energies. A rescaling of the fermionic fields in the context of 1-loop corrections has also been considered in \cite{Drukker:2011za}.

Given the above choice of $M^{\alpha}$ and $N^{\alpha}$, we find
\begin{empheq}{align}
	\textrm{det}\left(M^{\alpha}+N^{\alpha}Y^{\alpha}(R)\right)&=-i\left(\frac{A(\epsilon_0)}{A(R)}\right)^{\frac{1}{4}}I(R)\,.
\end{empheq}
Some algebra then shows that at large $R$
\begin{empheq}{align*}
	I(R)&\underset{R\rightarrow\infty}{\longrightarrow}\left\{
\begin{array}{ll}
	{\displaystyle\frac{e^{(\alpha E+\frac{1}{2})\left(R-\epsilon_0\right)}}{2\sqrt{1-e^{-2\epsilon_0}}\left(\alpha E+\frac{1}{2}\right)}\sqrt{\frac{1+\tanh\left(\epsilon_0+\sigma_0\right)}{2}}}\,, & \alpha E>-\frac{1}{2}
	\\\\
	{\displaystyle\frac{R}{\sqrt{1-e^{-2\epsilon_0}}}\sqrt{\frac{1+\tanh\left(\epsilon_0+\sigma_0\right)}{2}}}\,, & \alpha E=-\frac{1}{2}
	\\\\
	{\displaystyle-\frac{e^{-\left(\alpha E+\frac{1}{2}\right)\left(R-\epsilon_0\right)}}{2\sqrt{e^{2\epsilon_0}-1}\left(\alpha E-\frac{1}{2}\right)\left(\left(\alpha E-\frac{1}{2}\right)^2-1\right)}\sqrt{\frac{2}{1+\tanh\left(\epsilon_0+\sigma_0\right)}}}\;\times & 
	\\\\
	{\displaystyle\left(2\sinh\epsilon_0\left(\alpha E-\frac{1}{2}\right)^2-\frac{\cosh\left(2\epsilon_0+\sigma_0\right)}{\cosh\left(\epsilon_0+\sigma_0\right)}\left(\alpha E-\frac{1}{2}\right)+\frac{\sinh\sigma_0}{\cosh\left(\epsilon_0+\sigma_0\right)}\right)}\,, & \alpha E<-\frac{1}{2}
\end{array}
\right.\,.
\end{empheq}
The asymptotic expansion of the factor $A(\epsilon_0)/A(R)$ will not be necessary, as it will cancel out in the computations below.

We now deal with the projector $\mathcal{R}$ introduced in Sec. \ref{sec:GY}. The leading matrix coefficient in our case is $P_0=\gamma_{\underline{1}}$. Its two eigenvalues $\pm1$ fall on the real axis. Projection onto the subspace with eigenvalue $\pm1$ is achieved by acting with $\mathcal{R}_{\pm}=\frac{1}{2}\left(\mathds{1}\pm\gamma_{\underline{1}}\right)$.
We then find
\begin{empheq}{align}
	e^{\int_{\epsilon_0}^R\textrm{Tr}\left(\mathcal{R}_{\pm}P_1P_0^{-1}\right)}&=\left(\frac{A(R)}{A(\epsilon_0)}\right)^{\frac{1}{4}}e^{\pm i\alpha S}\,,
\end{empheq}
where
\begin{empheq}{align}
	S&=\int_{\epsilon_0}^Rd\sigma\,\frac{\sin^2\theta}{\sqrt{A}}\,.
\end{empheq}
Notice that this quantity is independent of $E$. The factor involving $A(R)/A(\epsilon_0)$ cancels against its inverse coming from $\textrm{det}\left(M^{\alpha}+N^{\alpha}Y^{\alpha}(R)\right)$ when introduced in the GY formula \eqref{eq: GY formula}. Moreover, the integral $S$ is finite in the $\epsilon_0\rightarrow0$ and $R\rightarrow\infty$ limits, as shown in App.~\ref{app: integral}, and its exponential contributes with a phase that depends on the charge of the fermions. Therefore, it will also cancel out once all the fermionic excitations are included. We shall omit it henceforth. 

Putting all the above results together and taking the ratio with the 1/2 BPS case given by $\sigma_0=\infty$, we find
\bea
	\Omega_E^{\alpha}(\sigma_0)&=&\left\{
\begin{array}{ll}
	{\displaystyle \ln\sqrt{\frac{1+\tanh\left(\epsilon_0+\sigma_0\right)}{2}}}\,, &\alpha  E\geq-\frac{1}{2}
	\\\\
	{\displaystyle-\ln\sqrt{\frac{1+\tanh\left(\epsilon_0+\sigma_0\right)}{2}}} &
	\\\\
	{\displaystyle +\ln\left(\frac{2\sinh\epsilon_0\left(\alpha E-\frac{1}{2}\right)^2-\frac{\cosh\left(2\epsilon_0+\sigma_0\right)}{\cosh\left(\epsilon_0+\sigma_0\right)}\left(\alpha E-\frac{1}{2}\right)+\frac{\sinh\sigma_0}{\cosh\left(\epsilon_0+\sigma_0\right)}}{2\sinh\epsilon_0\left(\alpha E-\frac{1}{2}\right)^2-e^{\epsilon_0}\left(\alpha E-\frac{1}{2}\right)+e^{-\epsilon_0}}\right)} \,,
	& \alpha E<-\frac{1}{2}
\end{array}
\right.\,.\cr &&
\label{fer-det}
\eea
These expressions are exact in $\epsilon_0$. 

\section{The $1$-loop effective action}
\label{sec:1loopaction}

After having found the 1-loop determinants for bosons and fermions, it is now time to sum the ratios $\Omega$'s over the energy label $E$. In this section, we explain in detail our summation procedure and derive the final result for the 1-loop effective action.

\subsection{Bosonic sums}

We start, as usual, by looking at the bosonic modes. As seen above, the operator ${\cal O}^{2,3,4}_E$ does not depend on $\sigma_0$ and consequently $\Omega_E^{2,3,4}=0$. The simplest non-trivial modes to consider are then the $\chi^{\underline{7},\underline{8},\underline{9}}$ modes, whose corresponding determinant is given in (\ref{Omega789}). That expression is symmetric with respect to $E=0$, which suggests regularizing the infinite sum over integer $E$ by a sharp cut-off $\Lambda\to\infty$, as follows 
\begin{empheq}{align}
	\sum_{E\,\in\,\mathds{Z}}\Omega^{7,8,9}_E&\longrightarrow\sum_{E=-\Lambda}^{\Lambda}\Omega^{7,8,9}_E\,.
\end{empheq}
Performing the sum, one readily obtains
\bea
	\sum_{E=-\Lambda}^{\Lambda}\Omega^{7,8,9}_E&=2\ln{\cal F}(\sigma_0,\Lambda)+\ln\tanh\sigma_0\,,
\label{sum789}
\eea
where
\be
{\cal F}(\sigma_0,\Lambda)\equiv\frac{\Gamma\left(\Lambda+1+\tanh \sigma_0\right)}{\Gamma\left(\Lambda+2\right)\Gamma\left(1+\tanh \sigma_0\right)}\,.
\ee	
The second term in (\ref{sum789}) is what will give the predicted result (\ref{prediction}) from the gauge theory. It comes from the $E=0$ mode of the $\Omega^{7,8,9}_E$ determinant.

It can be easily checked that the small $\epsilon_0$ and large $\Lambda$ limits commute for the bosonic determinants. In fact, we have already set $\epsilon_0=0$ in the result above. For large $\Lambda$, one obtains a logarithmic divergence
\be
\ln{\cal F}(\sigma_0,\Lambda)=\left(\tanh\sigma_0-1\right)\ln\Lambda-\ln\Gamma\left(1+\tanh\sigma_0\right)+{\cal O}(\Lambda^{-1})
\label{curlyFasymp}
\,,
\ee
which is going to cancel in the final result against similar contributions from the other modes. In fact, the full functions $\ln{\cal F}(\sigma_0,\Lambda)$ will cancel between the bosonic and fermionic sectors and the answer will be $\Lambda$-independent, as a consequence of supersymmetry. 

The next modes we consider are $\chi^{\underline{5},\underline{6}}$ with their determinant (\ref{Omega56}). In this case, we take as our starting point the formally symmetric, divergent sum
\begin{empheq}{align}
	\frac{1}{2}\sum_{E\,\in\,\mathds{Z}}\left(\Omega^{5,6}_E+\Omega^{5,6}_{-E}\right)\,.
	\label{sum56}
\end{empheq}
and regularize it by introducing an exponential suppression
\begin{empheq}{align}
	\frac{1}{2}\sum_{E\,\in\,\mathds{Z}}\left(\Omega^{5,6}_E+\Omega^{5,6}_{-E}\right)&\longrightarrow\frac{1}{2}\sum_{E\,\in\,\mathds{Z}}e^{-\mu|E|}\left(\Omega^{5,6}_E+\Omega^{5,6}_{-E}\right)\,.
	\label{sumreg56}
\end{empheq}
In the first term we shift $E\rightarrow E-1$ and in the second term we shift $E\rightarrow E+1$, as dictated by the multiplet structure (\ref{eq: multiplet 00}). Since each sum is now convergent, this is a perfectly legitimate operation. This can be understood as follows. To preserve supersymmetry at all steps of the computation, we want to sum over entire multiplets. Introducing a cut-off $\Lambda$, as we shall do presently, would break the multiplets at the extrema of the summing range, namely at $E=\pm \Lambda$, since for the fields $\chi^{\underline{5},\underline{6}}$ (and the fermions) the Fourier mode $E$ does not coincide with the $U(1)$ charge. In order to include all of the modes in a multiplet, we must make appropriate shifts. Of course, at large $\Lambda$ this becomes immaterial and all summing prescriptions (with our without shifts) gives the same asymptotic behavior. This procedure leads to
\bea
	\sum_{E\,\in\,\mathds{Z}}e^{-\mu|E|}\left(\Omega^{5,6}_E+\Omega^{5,6}_{-E}\right)
	&=&\sum_{E\,\in\,\mathds{Z}}e^{-\mu|E|}\left(\Omega^{5,6}_{E-1}+\Omega^{5,6}_{-E-1}\right)
	\cr
	&&+\sum_{E\,\in\,\mathds{Z}}\left(e^{-\mu|E-1|}-e^{-\mu|E|}\right)\Omega^{5,6}_{E-1}
	+\sum_{E\,\in\,\mathds{Z}}\left(e^{-\mu|E+1|}-e^{-\mu|E|}\right)\Omega^{5,6}_{-E-1}\,.\cr &&
\eea
The first line is still symmetric with respect to $E=0$, but the special mode is now located at $E=\pm1$. The second line is also symmetric under $E\rightarrow-E$, so we can write
\begin{empheq}{align}
	\frac{1}{2}\sum_{E\,\in\,\mathds{Z}}e^{-\mu|E|}\left(\Omega^{5,6}_E+\Omega^{5,6}_{-E}\right)
	=\frac{1}{2}\sum_{E\,\in\,\mathds{Z}}e^{-\mu|E|}\left(\Omega^{5,6}_{E-1}+\Omega^{5,6}_{-E-1}\right)
	+\mu\sum_{E=1}^{\infty}e^{-\mu E}\left(\Omega^{5,6}_{E-1}-\Omega^{5,6}_{-E-1}\right)\,,
\end{empheq}
up to terms that vanish for $\mu\rightarrow0$. The first sum will be divergent when we remove the regulator, but it can be regularized with a symmetric cutoff:
\begin{empheq}{align}
	\frac{1}{2}\sum_{E\,\in\,\mathds{Z}}e^{-\mu|E|}\left(\Omega^{5,6}_{E-1}+\Omega^{5,6}_{-E-1}\right)
	\longrightarrow
	\sum_{E=-\Lambda}^{\Lambda}\Omega^{5,6}_{E-1}=
	\ln{\cal F}(\sigma_0,\Lambda)+\ln\sqrt{\frac{1+\tanh\sigma_0}{2}}\,.
\end{empheq}
The second sum can be evaluated to give
\begin{empheq}{align}
	\mu\sum_{E=1}^{\infty}e^{-\mu E}\left(\Omega^{5,6}_{E-1}-\Omega^{5,6}_{-E-1}\right)&=-\ln\frac{1+\tanh\sigma_0}{2}\,,
\end{empheq}
in the $\mu\rightarrow0$ limit. Again we have set $\epsilon_0=0$ here. Putting everything together we find
\bea
	\frac{1}{2}\sum_{E\,\in\,\mathds{Z}}\left(\Omega^{5,6}_E+\Omega^{5,6}_{-E}\right)=
	\ln {\cal F}(\sigma_0,\Lambda) -\ln\sqrt{\frac{1+\tanh\sigma_0}{2}}
	\,,
	\label{alt1}
\eea
The second term in this result is ultimately responsible for the disagreement between the gauge theory prediction and the string theory calculation.

\subsection{Fermionic sums} 

For the fermionic modes we start with the $\mu$-regularized sum as done above for $\Omega^{5,6}_E$, with $E$ being now summed over half-integer values:
\begin{empheq}{align}
	\frac{1}{2}\sum_{E\,\in\,\mathds{Z}+\frac{1}{2}}\left(\Omega^\alpha_E+\Omega^\alpha_{-E}\right)&\longrightarrow\frac{1}{2}\sum_{E\,\in\,\mathds{Z}+\frac{1}{2}}e^{-\mu|E|}\left(\Omega^\alpha_E+\Omega^\alpha_{-E}\right)\,.
\end{empheq}
We make the shifts $E\rightarrow E+\frac{\alpha}{2}$ in the first term and $E\rightarrow E-\frac{\alpha}{2}$ in the second. The resulting sums are over integer energies.  These shifts are motivated, again, by the supermultiplet structure (\ref{eq: multiplet 00}). In the small $\mu$ limit, one finds
\bea
\frac{1}{2}\sum_{E\,\in\,\mathds{Z}}e^{-\mu|E|}\left(\Omega^\alpha_{E+\frac{\alpha}{2}}+\Omega^\alpha_{-E+\frac{\alpha}{2}}\right)
-\frac{\alpha\mu}{2}\sum_{E=1}^{\infty}e^{-\mu E}\left(\Omega^\alpha_{E+\frac{\alpha}{2}}-\Omega^\alpha_{-E+\frac{\alpha}{2}}\right)\,.
	\label{sumintferm}
\eea
To compute the first sum, we introduce a symmetric cutoff:
\begin{empheq}{align}
	\frac{1}{2}\sum_{E\,\in\,\mathds{Z}}e^{-\mu|E|}\left(\Omega^\alpha_{E+\frac{\alpha}{2}}+\Omega^\alpha_{-E+\frac{\alpha}{2}}\right)&\longrightarrow \frac{1}{2}\sum_{E=-\Lambda}^{\Lambda}\left(\Omega^\alpha_{E+\frac{\alpha}{2}}+\Omega^\alpha_{-E+\frac{\alpha}{2}}\right)\,.
\end{empheq}

At this point we encounter a difference with respect to the bosonic case, in which taking $\epsilon_0$ small and summing over $-\Lambda\le E\le \Lambda$ to then send $\Lambda$ to infinity were two commuting operations. For the fermions this is no longer the case. Summing over the energies and taking $\Lambda$ large before sending $\epsilon_0$ to zero produces a logarithmic divergence in $\epsilon_0$, as well as a logarithmic divergence in $\Lambda$ that does not cancel, in the final result, against the similar divergences coming from the bosonic sector. This is explained in detail in App.~\ref{app-div}. We believe these surviving divergences to not have a physical interpretation, being probably due to an artifact of the regularization procedure. Notice in fact that both $\Lambda$ and $\epsilon_0$ are large energy cut-offs, so that this regularization is somehow redundant. We leave a deeper understanding of this issue for the future. Here we take the small $\epsilon_0$ limit before summing over energies. As a result, the  fermionic determinant (\ref{fer-det}) reduces to
\begin{eqnarray}
\Omega_E^{\alpha}&=&
\left\{
\begin{array}{ll}
	{\displaystyle \ln\sqrt{\frac{1+\tanh\sigma_0}{2}}} \,, \qquad &\alpha  E\geq-\frac{1}{2}
	\\
	 & 
	 \\
	{\displaystyle-\ln\sqrt{\frac{1+\tanh\sigma_0}{2}}}+
	{\displaystyle\ln\frac{\left(\alpha E-\frac{1}{2}\right)-\tanh\sigma_0}{\left(\alpha E-\frac{1}{2}\right)-1}}\,, \qquad & 
	\alpha E<-\frac{1}{2}
\end{array}
\right.\,.
\end{eqnarray}
Using this expression for the $\alpha=1$ case, we see that the first sum in (\ref{sumintferm}) evaluates to
\be
\frac{1}{2}\sum_{E=-\Lambda}^{\Lambda}\left(\Omega^\alpha_{E+\frac{\alpha}{2}}+\Omega^\alpha_{-E+\frac{\alpha}{2}}\right)=
\ln{\cal F}(\sigma_0,\Lambda)+
\ln\sqrt{\frac{1+\tanh\sigma_0}{2}}\,,
\ee
whereas the second sum in the limit of $\mu\to\infty$ gives
\be
-\frac{\mu}{2}\sum_{E=1}^{\infty}e^{-\mu E}\left(\Omega^+_{E+\frac{1}{2}}-\Omega^+_{-E+\frac{1}{2}}\right)=
-\ln\sqrt{\frac{1+\tanh\sigma_0}{2}}\,.
\ee
The final result for the $\alpha=+1$ fermions is therefore
\begin{eqnarray}
	&& \frac{1}{2}\sum_{E\,\in\,\mathds{Z}+\frac{1}{2}}e^{-\mu|E|}\left(\Omega^+_E+\Omega^+_{-E}\right)=
	\ln{\cal F}(\sigma_0,\Lambda)
\label{sumfinferm}
\,.
\end{eqnarray}
The case $\alpha=-1$ yields exactly the same result.

\subsection{Final result}

We have now all the ingredients to evaluate the difference (\ref{final-dif}) between the 1-loop effective actions of the 1/4 BPS and the 1/2 BPS string configurations. Using (\ref{O234}), (\ref{sum789}), (\ref{alt1}), and (\ref{sumfinferm}), we find
\bea
	\Delta\Gamma^{1-\textrm{loop}}_{\textrm{effective}}&=&\frac{3}{2}\ln\tanh\sigma_0-\ln\sqrt{\frac{1+\tanh\sigma_0}{2}}
	=
	\frac{3}{2}\ln\cos\theta_0-\ln\cos\frac{\theta_0}{2}
	\,,
\eea
where in the last equality we have used the relation (\ref{sigma0theta0}) between $\sigma_0$ and $\theta_0$. Notice that the $\ln{\cal F}(\sigma_0,\lambda)$ terms (and with them the $\Lambda$ dependence) cancel exactly between the bosonic and fermionic sectors, even before taking the large $\Lambda$ limit. This is a consequence of supersymmetry. Had we not shifted the energies in the sums over the 5 and 6 modes and the fermions, this cancellation would have taken place only asymptotically for large $\Lambda$.

Since $\vev{W}\simeq e^{-\Gamma_\textrm{effective}}$, we see that we find a result which differs from the gauge theory prediction (\ref{prediction}) by the finite discrepancy $\ln\cos\frac{\theta_0}{2}$. This is the same discrepancy that has recently been found, using a numerical procedure, in \cite{Forini:2015bgo}. 

An important observation, on which we shall return later on, is that we are able to track the origin both of the predicted term and of the discrepancy. The former originates from the special modes, $E=0$, of the $\Omega^{7,8,9}_E$ determinant (\ref{Omega789}), whereas the latter comes from the $\Omega^{5,6}_E$ determinant (\ref{Omega56}). More specifically, the discrepancy could be removed, if we were to modify {\it ad hoc} the sum over $\Omega^{5,6}$ as follows
\be
\sum_{E\in \mathbb{Z}}\Omega^{5,6}_E \longrightarrow \sum_{E=-\Lambda-1}^\Lambda \Omega^{5,6}_E\,.
\ee
Unfortunately, there does not seem to be a justification for this summing prescription and, therefore, we discard this possibility.

\section{Conclusions}
\label{Sec:Conclusions}

In this paper we have computed the 1-loop effective action of quantum string fluctuations around the classical string configuration dual to the 1/4 BPS latitude Wilson loop. More specifically, we have considered the ratio between the 1/4 BPS string configuration and the configuration associated to the 1/2 BPS circular loop. The rationale for this course of action was to remove possible sources of ambiguity related to string ghost zero modes, which are supposed to cancel between two string configurations with the same world-sheet topology, as originally argued in \cite{Drukker:2000rr} and later proposed in \cite{Kruczenski:2008zk}. Our final objective was to match this string theory computation to the gauge theory prediction (\ref{prediction}) obtained via supersymmetric localization.

We have paid close attention to the group theoretical structure of the fluctuations, which are neatly organized in supermultiplets of the $SU(2|2)$ supergroup preserved by the latitude. This supermultiplet organization has consequences in the way the sums over energies have to be performed.  
One salient feature of our computation is that it is fully analytical. Technically, the result relied on our choice to work with the linear fermionic operator, rather than with the square of it, as customarily done in the literature. 

Unfortunately, we have not found agreement between the gauge theory prediction and the string theory result. We have found instead a finite discrepancy that has also been reported recently in \cite{Forini:2015bgo}, having been obtained there using a different procedure than ours. Barring a simple oversight in our work or in \cite{Forini:2015bgo}, there are several possible reasons for the disagreement which are worth exploring, either in string theory, where there might still be an unresolved subtlety in the procedure for computing the determinants, or in the gauge theory prediction. In this regard, let us mention that there exists another claim of disagreement in the subleading order at strong coupling, this time in the computation of correlators of latitudes \cite{Bassetto:2009ms}.\footnote{We thank L. Griguolo for reminding us of this previous result.} 

Despite the disagreement, we think we have learned something from this computation. Specifically, one observation that we find intriguing is the fact that we could track the origin of the correct, expected result to some very specific mode: the $E=0$ mode of the $\Omega^{7,8,9}_E$ determinant associated to the fields charged under the $SU(2)_B$ factor of the supergroup preserved by the 1/4 BPS latitude. This observation, of course, begs the question of whether this might be a more general phenomenon. If this is confirmed to be true for other Wilson loops ({\it e.g.}, the DGRT loops of \cite{Drukker:2007qr}), perhaps it might hint at the existence of some `dual' localization mechanism in string theory, in which the string partition function is captured entirely by some special modes, in the same way in which, on the gauge theory side, the operator's expectation value is captured by the zero modes of a scalar field \cite{Pestun:2007rz}. Of course, this by itself would not solve the puzzle of the presence of a discrepancy, that should be better understood and eventually eliminated, but it points to an interesting direction worth exploring.

The structure of our result and the explicit cancellations that we have displayed shine a ray of hope in the prospect of bulk localization with extended objects. In fact, there has recently been some effort in reproducing the full exact results of localization from physics in the bulk. One natural ingredient in this attempt would be an off-shell formulation of the bulk theory. For example, in the attempt to obtain the full ABJM partition function from gravity \cite{Dabholkar:2014wpa}, the off-shell theory was provided by conformal supergravity. A related result was also the match between partition functions beyond leading order obtained in \cite{Bhattacharyya:2012ye}. Interestingly, in \cite{Bhattacharyya:2012ye} the full 1-loop result originates from a zero mode present on the 11-dimensional supergravity side, similarly to what happens in our setting. To an optimistic reader this points to a potential bulk localization circumventing the need for an off-shell string action. This statement is highly speculative but certainly worth checking in other related setups, where on the holographic side strings and branes are involved. We hope to report soon on further tests of this idea.

To conclude, we believe to be worthwhile to attempt high precision tests of the AdS/CFT correspondence, as the one presented here. Given the plethora of exact results obtained via localization in supersymmetric field theories with gravity duals, it is important to reproduce those results in string theory. One of the explicit benefits of such attempts will undoubtedly be a better understanding of string perturbation theory in curved spaces beyond the semiclassical approximation. 

\section*{Acknowledgments}

We are thankful to O. Aharony, N. Berkovits, A. Dabholkar, L. Griguolo, Z. Komargodski, M. Kruczenski, W. M\"uck, D. Seminara, and E. Vescovi for discussions and correspondence. AF is supported by CONICYT/PAI ``Apoyo al Retorno'' grant 821320022. GAS is thankful to the ICTP Associate Program for hospitality during the gestation phase of this project and to Pict 2012-0417 for financial support. LAPZ acknowledges the hospitality of CINVESTAV, Mexico, the University of Naples {\it Federico II}, Italy and Tel Aviv University, Israel. He also thanks J. Sonnenschein for encouragement. DT acknowledges PUC Chile for hospitality during this project and CNPq and FAPESP (grants 2014/18634-9 and 2015/17885-0) for partial financial support.

\appendix
\section{Geometric data}
\label{App: geometric data}

In this appendix we collect all the relevant geometric quantities for the calculation of the spectrum of the string fluctuations. Target space indices are denoted by $m,n,\ldots$, worldvolume indices are $a,b,\ldots$, directions orthogonal to the string are represented by $i,j,\ldots$. All corresponding tangent space indices are underlined.

We start by constructing an adapted $EAdS_5\times S^5$ vielbein $E^{\underline{m}}=\left(E^{\underline{a}},E^{\underline{i}}\right)$. For the case at hand, the simplest choice is
\begin{empheq}{align}
	& E^{\underline{0}}=\frac{\cosh^2u\,\sinh^2\rho\,d\psi+\sin^2\theta\,d\phi}{\sqrt{A(u,\rho,\theta)}}\,,
	\qquad
	E^{\underline{1}}=\frac{\cosh^2u\,\rho'\,d\rho+\theta'\,d\theta}{\sqrt{B(u,\rho,\theta)}}\,,
	\cr
	& E^{\underline{2}}=\sinh u\,d\vartheta\,,\qquad
	E^{\underline{3}}=\sinh u\,\sin\vartheta\,d\varphi\,,\qquad
	E^{\underline{4}}=du\,,\cr
	& E^{\underline{5}}=\frac{\cosh u\left(\rho'\,d\theta-\theta'\,d\rho\right)}{\sqrt{B(u,\rho,\theta)}}\,,
	\qquad
	E^{\underline{6}}=\frac{\cosh u\,\sinh\rho\sin\theta\left(d\phi-d\psi\right)}{\sqrt{A(u,\rho,\theta)}}\,,
	\cr
	& E^{\underline{7}}=\cos\theta\,d\xi\,,
	\qquad
	E^{\underline{8}}=\cos\theta\cos\xi\,d\alpha_1\,,
	\qquad
	E^{\underline{9}}=\cos\theta\sin\xi\,d\alpha_2\,,
\end{empheq}
where
\begin{empheq}{align}
	A(u,\rho,\theta)=\cosh^2u\,\sinh^2\rho+\sin^2\theta\,,
	\qquad
	B(u,\rho,\theta)=\cosh^2u\,\rho'^2+\theta'^2\,,
\end{empheq}
and $\rho'$ and $\theta'$ are understood as functions of $\rho$ and $\theta$, respectively, {\it e.g.} $\rho'=-\sinh\rho$ and $\theta'=-\sin\theta$. To allow for a more general gauge, we will consider the rotation
\begin{empheq}{align}
	\left(
	\begin{array}{c}
		E^{\underline{5}}
		\\
		E^{\underline{6}}
	\end{array}
	\right)
	&\longrightarrow
	\left(
	\begin{array}{cc}
		\cos\delta(\psi,\phi) & \sin\delta(\psi,\phi)
		\\
		-\sin\delta(\psi,\phi) & \cos\delta(\psi,\phi)
	\end{array}
	\right)
	\left(
	\begin{array}{c}
		E^{\underline{5}}
		\\
		E^{\underline{6}}
	\end{array}
	\right)\,,
\end{empheq}
where $\delta(\psi,\phi)$ is an arbitrary function to be fixed at our convenience. As advertised in the main text, upon taking the pullback onto the worldvolume, the first two components give a vielbein for the induced geometry, namely,
\begin{empheq}{alignat=5}
	e^{\underline{0}}&\equiv P[E^{\underline{0}}]&&=\sqrt{A}\,d\tau\,,
	&\qquad
	e^{\underline{1}}&\equiv P[E^{\underline{1}}]&&=\sqrt{A}\,d\sigma\,,
\end{empheq}
while the remaining components vanish. The conformal factor reads
\begin{empheq}{align}
	A=\sinh^2\rho+\sin^2\theta
	=\frac{4\cosh\left(2\sigma+\sigma_0\right)\cosh\sigma_0}{\left(\sinh\left(2\sigma+\sigma_0\right)-\sinh\sigma_0\right)^2}\,.
\end{empheq}
The pullback of the target space spin connection is
\begin{empheq}{alignat=5}
	w&\equiv P\left[\Omega^{\underline{01}}\right]&&=-\frac{\sinh^2\rho\cosh\rho+\sin^2\theta\cos\theta}{A}\,d\tau
	=\frac{A'}{2A}\,d\tau\,,
	\\
	\mathcal{A}&\equiv P\left[\Omega^{\underline{56}}\right]&&=\frac{\sinh^2\rho\cos\theta+\cosh\rho\sin^2\theta}{A}\,d\tau-d\delta
=\tanh\left(2\sigma+\sigma_0\right)\,d\tau-d\delta\,,
\end{empheq}
corresponding, respectively, to the induced geometry's spin connection and a $U(1)$ connection in the normal bundle that gauges rotations in the $5$-$6$ plane. We also have
\begin{empheq}{alignat=5}
	P\left[\Omega^{\underline{05}}\right]&=-P\left[\Omega^{\underline{16}}\right]\,=m\left(\cos\delta\,d\tau-\sin\delta\,d\sigma\right)\,,
	\\
	P\left[\Omega^{\underline{06}}\right]&=P\left[\Omega^{\underline{15}}\right]\,=-m\left(\sin\delta\,d\tau+\cos\delta\,d\sigma\right)\,,
\end{empheq}
where
\begin{empheq}{align}
	m=\frac{\sinh\rho\sin\theta\left(\cosh\rho-\cos\theta\right)}{A}
	=\frac{1}{\cosh\left(2\sigma+\sigma_0\right)}\,.
\end{empheq}
From the relation $H^{\underline{i}}_{\phantom{\underline{i}}ab}=P[\Omega^{\underline{i}}_{\phantom{\underline{i}}\underline{a}}]_ae^{\underline{a}}_{\phantom{\underline{a}}b}$, 
we read the extrinsic curvatures of the embedding:
\begin{empheq}{alignat=5}
	H^{\underline{5}\phantom{a}b}_{\phantom{\underline{5}}a}&=\frac{m}{\sqrt{A}}\left(
	\begin{array}{cc}
		-\cos\delta & \sin\delta \\
		\sin\delta & \cos\delta
	\end{array}
	\right)\,,
	&\qquad
	H^{\underline{6}\phantom{a}b}_{\phantom{\underline{6}}a}&=\frac{m}{\sqrt{A}}\left(
	\begin{array}{cc}
		\sin\delta & \cos\delta \\
		\cos\delta & -\sin\delta
	\end{array}
	\right)\,.
\end{empheq}
These tensors are traceless as a consequence of the equations of motion $g^{ab}H^{\underline{i}}_{\phantom{\underline{i}}ab}=0$. We will sometimes abuse notation and call $w=w_{\tau}$ and $\mathcal{A}=\mathcal{A}_{\tau}$. Notice that
\begin{empheq}{alignat=5}
	w-\mathcal{A}=-\left(\cosh\rho+\cos\theta\right)+\partial_{\tau}\delta\,,
	\qquad	
	m^2=\frac{1}{2}\partial_{\sigma}\mathcal{A}\,,
	\qquad
	\partial_{\sigma}w-\partial_{\sigma}\mathcal{A}=\sinh^2\rho-\sin^2\theta\,.
	\label{wminusA}
\end{empheq}

Another piece of information we need involves contractions of the Riemann curvature of the form $\delta^{\underline{ab}}R_{\underline{aibj}}$. We find that the only non-vanishing components are
\begin{empheq}{alignat=5}
	&\delta^{\underline{ab}}R_{\underline{a2b2}}=\delta^{\underline{ab}}R_{\underline{a3b3}}=\delta^{\underline{ab}}R_{\underline{a4b4}}=-\frac{2\sinh^2\rho}{A}\,,\cr
	&\delta^{\underline{ab}}R_{\underline{a7b7}}=\delta^{\underline{ab}}R_{\underline{a8b8}}=\delta^{\underline{ab}}R_{\underline{a9b9}}=\frac{2\sin^2\theta}{A}\,.
\end{empheq}

It remains to look at the contribution from the RR field strength to the spinor covariant derivative. In terms of tangent components we have, for $\delta(\psi,\phi)=0$,
\bea
	\textrm{vol}\left(EAdS_5\right)&=&-\frac{1}{\sqrt{AB}}\left(\cosh^2(u)\sinh\rho\,\rho'E^{\underline{0}}\wedge E^{\underline{1}}-\cosh(u)\sinh\rho\,\theta'E^{\underline{0}}\wedge E^{\underline{5}}\right.
	\cr
	&&\hskip1.5cm \left.+\cosh(u)\sin\theta\,\rho'E^{\underline{1}}\wedge E^{\underline{6}}-\sin\theta\,\theta'E^{\underline{5}}\wedge E^{\underline{6}}\right)\wedge E^{\underline{2}}\wedge E^{\underline{3}}\wedge E^{\underline{4}}\,,\cr
	 \slashed{F}_5&=&\frac{4i}{\sqrt{AB}}\left(\sinh\rho\,\rho'\Gamma^{\underline{01}}-\sinh\rho\,\theta'\Gamma^{\underline{05}}+\sin\theta\,\rho'\Gamma^{\underline{16}}-\sin\theta\,\theta'\Gamma^{\underline{56}}\right)\Gamma^{\underline{234}}\left(1-\Gamma^{11}\right)\,.\cr&&
\eea
The expression that actually enters in the fermionic action is
\begin{empheq}{align}
g^{ab}\Gamma_a\slashed{F}_5\Gamma_b&=\frac{8i}{\sqrt{AB}}\left(\sinh\rho\,\rho'\Gamma^{\underline{01}}+\sin\theta\,\theta'\Gamma^{\underline{56}}\right)\Gamma^{\underline{234}}\left(1+\Gamma^{11}\right)\,.
\end{empheq}
Notice that $\Gamma^{\underline{56}}$ is invariant under rotations in the $5$-$6$ plane, so this is actually valid for any $\delta(\psi,\phi)$.


\section{Dimensional reduction of spinors}
\label{App: Dimensional reduction}

Given the symmetries of our problem, the natural way to decompose the 10-dimensional Lorentz group (in Lorentzian signature) is
\begin{empheq}{align}
	 SO(9,1)&\supset\underbrace{SO(2)}_{\gamma}\times\underbrace{SO(2,1)}_{\rho}\times\underbrace{SO(2)}_{\tau}\times\underbrace{SO(3)}_{\lambda}\,,
\end{empheq}
corresponding to the $(0,1)$, $(2,3,4)$, $(5,6)$ and $(7,8,9)$ tangent directions, respectively. Under this decomposition, a possible representation of the 10-dimensional gamma matrices is
\begin{empheq}{alignat=5}
	 \Gamma_{\underline{a}}&=\gamma_{\underline{a}}\otimes\mathds{1}\otimes\mathds{1}\otimes\mathds{1}\otimes\sigma_1\,,
	&\qquad
	\underline{a}&=0,\,1\,,
	\cr
	 \Gamma_{\underline{i}}&=\left(-i\gamma_{\underline{01}}\right)\otimes\rho_{\underline{i}}\otimes\mathds{1}\otimes\mathds{1}\otimes\sigma_1\,,
	&\qquad
	\underline{i}&=2,\,3,\,4\,,
	\cr
	\Gamma_{\underline{i}}&=\mathds{1}\otimes\mathds{1}\otimes\tau_{\underline{i}}\otimes\mathds{1}\otimes\sigma_2\,,
	&\qquad
	\underline{i}&=5,\,6\,,
	\cr
	\Gamma_{\underline{i}}&=\mathds{1}\otimes\mathds{1}\otimes \left(-i\tau_{\underline{56}}\right)\otimes\lambda_{\underline{i}}\otimes\sigma_2\,,
	&\qquad
	\underline{i}&=7,\,8,\,9\,,
\end{empheq}
where we named the Dirac matrices associated to each factor as displayed above. We also choose the $SO(2,1)$ and $SO(3)$ Clifford algebra representations where $\rho_{\underline{234}}=1$ and $\lambda_{\underline{789}}=i.$\footnote{Recall that in odd dimensions there are two inequivalent representations of the Clifford algebra that differ by the value of the would-be chirality matrix.} The chirality matrix is then
\begin{empheq}{align}
	\Gamma^{11}\equiv\Gamma_{\underline{0123456789}}
	=\mathds{1}\otimes\mathds{1}\otimes\mathds{1}\otimes\mathds{1}\otimes\sigma_3\,.
\end{empheq}

For the intertwiners, which specify the conjugation properties of the gamma matrices, we have\footnote{The charge conjugation matrix is related to $B$ by $C=B^TA$, where $A$ is the matrix used to define the Dirac conjugate $\overline{\psi}=\psi^{\dagger}A$.}
\begin{empheq}{alignat=5}
	B_{(2,0)\pm}\gamma_{\underline{a}}B^{-1}_{(2,0)\pm}&=\pm\gamma_{\underline{a}}^*\,,
	&\qquad
	B^{\dagger}_{(2,0)\pm}B_{(2,0)\pm}&=\mathds{1}\,,
	&\qquad
	\underline{a}&=0,\,1\,,
\end{empheq}
and
\begin{empheq}{alignat=5}
	B_{(2,0)\pm}\tau_{\underline{i}}B^{-1}_{(2,0)\pm}&=\pm\tau_{\underline{i}}^*\,,
	&\qquad
	B^{\dagger}_{(2,0)\pm}B_{(2,0)\pm}&=\mathds{1}\,,
	&\qquad
	\underline{i}&=5,\,6\,,
\end{empheq}
for the $SO(2)$ factors,
\begin{empheq}{alignat=5}
	B_{(3,0)}\lambda_{\underline{i}}B^{-1}_{(3,0)}&=-\lambda_{\underline{i}}^*\,,
	&\qquad
	B^{\dagger}_{(3,0)}B_{(3,0)}&=\mathds{1}\,,
	&\qquad
	\underline{i}&=7,\,8,\,9\,,
\end{empheq}
for $SO(3)$, and
\begin{empheq}{alignat=5}
	B_{(2,1)}\rho_{\underline{a}}B^{-1}_{(2,1)}&=\rho_{\underline{a}}^*\,,
	&\qquad
	B^{\dagger}_{(2,1)}B_{(2,1)}&=\mathds{1}\,,
	&\qquad
	\underline{i}&=2,\,3,\,4\,,
\end{empheq}
for $SO(2,1)$. With this information we can build
\bea
	B_{(9,1)+}&=&B_{(2,0)-}\otimes B_{(2,1)}\otimes B_{(2,0)+}\otimes B_{(3,0)}\otimes\sigma_3\,,
	\cr
	B_{(9,1)-}&=&B_{(2,0)-}\otimes B_{(2,1)}\otimes B_{(2,0)+}\otimes B_{(3,0)}\otimes\mathds{1}\,,
\eea
which satisfy
\begin{empheq}{alignat=5}
	B_{(9,1)\pm}\Gamma_{\underline{m}}B^{-1}_{(9,1)\pm}&=\pm\Gamma_{\underline{m}}^*\,,
	&\qquad
	B^{\dagger}_{(9,1)\pm}B_{(9,1)\pm}&=\mathds{1}\,,
	&\qquad
	\underline{m}&=0,\,1,\,\ldots,\,9\,.
\end{empheq}

To dimensionally reduce the type IIB spinor $\theta$, we start by looking at the Weyl condition. We see that a 10-dimensional positive chirality spinor has the form
\begin{empheq}{align}
	\theta&=\theta_{(2,0)}\otimes\theta_{(2,1)}\otimes\theta'_{(2,0)}\otimes\theta_{(3,0)}\otimes\left(
 \begin{array}{c}
  1 \\ 0
   \end{array}
    \right)\,.
\end{empheq}
The Majorana condition, which reads
$\theta^*=B_{(9,1)+}\theta$,
implies
\begin{empheq}{align}
	 \theta_{(2,0)}^*\otimes\theta_{(2,1)}^*\otimes{\theta'_{(2,0)}}^*\otimes\theta_{(3,0)}^*&=B_{(2,0)-}\theta_{(2,0)}\otimes B_{(2,1)}\theta_{(2,1)}\otimes B_{(2,0)+}\theta'_{(2,0)}\otimes B_{(3,0)}\theta_{(3,0)}\,.
\end{empheq}
In $2+0$ and $2+1$ dimensions Majorana spinors exist, but there are no pseudo-Majorana spinors in $2+0$. Moreover, in $3+0$ dimensions there are no possible reality constraints on a single spinor. It will prove convenient to introduce two $SO(2)$ basis spinors $\eta_{\pm}$ satisfying\footnote{For example, in the representation $\tau_{\underline{5}}=\sigma_1$, $\tau_{\underline{6}}=\sigma_2$ we have $B_{(2,0)+}=\sigma_1$ with $\eta_+=\left(\begin{array}{c} 1 \\ 0 \end{array}\right)$ and $\eta_-=\left(\begin{array}{c} 0 \\ 1 \end{array}\right)$.}
\begin{empheq}{align}
	\eta_{\alpha}^*&=B_{(2,0)+}\eta_{-\alpha}\,,
\end{empheq}
as well as two $SO(3)$ basis spinors $\zeta_{\pm}$ with\footnote{In the representation $\lambda_{\underline{7}}=\sigma_1$, $\lambda_{\underline{8}}=\sigma_2$ and $\lambda_{\underline{9}}=\sigma_3$ we have $B_{(3,0)}=\sigma_2$ with $\zeta_+=\left(\begin{array}{c} 1 \\ 0 \end{array}\right)$ and $\zeta_-=\left(\begin{array}{c} 0 \\ 1 \end{array}\right)$.}
\begin{empheq}{align}
	\zeta_{\alpha}^*&=i\alpha B_{(3,0)}\zeta_{-\alpha}\,.
\end{empheq}
For $SO(2,1)$ we can introduce\footnote{In the representation $\rho_{\underline{2}}=\sigma_1$, $\rho_{\underline{3}}=\sigma_3$ and $\rho_{\underline{4}}=i\sigma_2$ we have $B_{(2,1)}=\mathds{1}$ with $\chi_+=\left(\begin{array}{c} 1 \\ 0 \end{array}\right)$ and $\chi_-=\left(\begin{array}{c} 0 \\ 1 \end{array}\right)$.}
\begin{empheq}{align}
	\chi_{\alpha}^*&=B_{(2,1)}\chi_{\alpha}\,.
\end{empheq}
We can then write
\begin{empheq}{align}
	 \theta&=\sum_{\alpha,\alpha',\alpha''=\pm}\psi^{\alpha'\alpha''}_{\alpha}\otimes\chi_{\alpha}\otimes\eta_{\alpha''}\otimes\zeta_{\alpha'}\otimes\left( \begin{array}{c}
	 1 \\
	  0
	   \end{array}
	    \right)\,.
\end{empheq}
The reality constraints imply ${\psi^{\alpha'\alpha''}_{\alpha}}^*=i\alpha'' B_{(2,0)-}\psi^{\alpha'-\alpha''}_{-\alpha}$.

The 2-dimensional spinors $\psi^{\alpha'\alpha''}_{\alpha}$ transform in the $\left(\square,\square\right)$ representations of
$su(2)\times su(2)\simeq so(2,1)\times so(3)$ and have $U(1)\simeq SO(2)$ charge $\alpha/2$. The total number of real
degrees of freedom is 16, as appropriate. We can choose to represent them by the four Dirac spinors
\begin{empheq}{alignat=5}
	\psi^{++}_+,&\quad\psi^{--}_+,&\quad\psi^{+-}_+,&\quad\psi^{-+}_+\,,
\end{empheq}
all of which have charge $1/2$. In the Euclidean continuation the Majorana condition is lost and we end up with 8 independent Dirac spinors.

\section{The integral $S$}
\label{app: integral}

In Sec.~\ref{Sec:Determinants} we encountered the integral
\begin{empheq}{align}
	S&=\int_{\epsilon_0}^R d\sigma\,\frac{\sin^2\theta}{\sqrt{A}}\,.
	\label{intS}
\end{empheq}
Here we will compute it explicitly and show that it is finite in the limits $R\rightarrow\infty$ and $\epsilon_0\rightarrow0$. To this purpose, let us first notice that
\begin{empheq}{align}
	d\sigma\,\frac{\sin^2\theta}{\sqrt{A}}&=-d\theta\,\frac{\left(\cos\theta-\cos\theta_0\right)}{\sqrt{\left(\cos\theta-\cos\theta_0\right)^2+\sin^2\theta_0}}\,.
\end{empheq}
The substitution
\begin{empheq}{alignat=5}
	\theta&=2\arctan\left(uu_0\right)\,,
	&\qquad
	u_0&=\sqrt{\tan\frac{\theta_0}{2}}
\end{empheq}
gives
\begin{empheq}{align}
	-d\theta\,\frac{\left(\cos\theta-\cos\theta_0\right)}{\sqrt{\left(\cos\theta-\cos\theta_0\right)^2+\sin^2\theta_0}}&=du\,\frac{2u_0}{\sqrt{1+u_0^4}}\frac{\left(u^2-u_0^2\right)}{\left(1+u_0^2u^2\right)\sqrt{1+u^4}}\,,
\end{empheq}
which can be integrated in terms of incomplete elliptic integrals of the first and third kind
\begin{empheq}{align}
	F(z|k)=\int_0^z\frac{du}{\sqrt{1-u^2}\sqrt{1-k^2u^2}}\,,
	\qquad
	\Pi(z;\nu|k)=\int_0^z\frac{du}{(1-\nu u^2)\sqrt{1-u^2}\sqrt{1-k^2u^2}}\,.
\end{empheq}
We find
\begin{empheq}{align}
	\int d\sigma\,\frac{\sin^2\theta}{\sqrt{A}}&=\frac{2e^{-\frac{i\pi}{4}}}{u_0\sqrt{1+u_0^4}}\left[F\left(e^{\frac{i\pi}{4}}u\big|i\right)-\left(1+u_0^4\right)\Pi\left(e^{\frac{i\pi}{4}}u,e^{\frac{i\pi}{2}}u_0^2\big|i\right)\right]\,.
\end{empheq}
The upper and lower limits of integration in (\ref{intS}) are mapped, respectively, to $u\to 0$ and $u\to u_0$. By definition, the elliptic integrals vanish at $z=0$. Thus,
\begin{empheq}{align}
	\int_0^{\infty}d\sigma\,\frac{\sin^2\theta}{\sqrt{A}}&=-\frac{2e^{-\frac{i\pi}{4}}}{u_0\sqrt{1+u_0^4}}\left[F\left(e^{\frac{i\pi}{4}}u_0\big|i\right)-\left(1+u_0^4\right)\Pi\left(e^{\frac{i\pi}{4}}u_0,e^{\frac{i\pi}{2}}u_0^2\big|i\right)\right]\,.
\end{empheq}

\section{Swapping the $\epsilon_0\to0$ and $\Lambda\to\infty$ limits}
\label{app-div}

In this appendix, we compute the sum (\ref{sumintferm}) without taking the small $\epsilon_0$ limit first, namely using the full expression (\ref{fer-det}) for the fermionic determinant. The crucial difference is that, inside of the logarithms, there are now quadratic terms in $E$ with coefficients that vanish as $\epsilon_0$ goes to zero. 

It is going to be convenient to use the following results
\begin{eqnarray}
&&\hskip -.7cm
\sum_{E=1}^\Lambda \ln\left(a E^2+b E+c\right)=
(2\ln\Lambda+\ln a-2)\Lambda+\left(1+\frac{b}{a}\right)\ln\Lambda
+ \ln
\frac{2(a+b+c)\pi}{a\Gamma(\Delta_+)\Gamma(\Delta_-)}
+{\cal O}(\Lambda^{-1/2})\,,
\cr &&
\label{sum_math}
\end{eqnarray}
with $\Delta_\pm=\frac{4a+b\pm\sqrt{b^2-4ac}}{2a}$, and
\begin{eqnarray}
&& \frac{\alpha\mu}{2}\sum_{E=1}^\infty
e^{-\mu E} \ln\frac{a E^2+ b E+c }{a E^2 +b_0 E+c_0}={\cal O}(\mu)\,.
\label{secform}
\end{eqnarray}
We focus on the case $\alpha=1$. The first term of the sum (\ref{sumintferm}) contains
\begin{eqnarray}
\sum_{E=1}^\Lambda \left(\Omega^+_{E+\frac{1}{2}}+\Omega^+_{-E+\frac{1}{2}}\right)&=&
\sum_{E=1}^\Lambda \ln\frac{a E^2+bE+c}{a E^2+b_0 E+c_0}\,,
\end{eqnarray}
with
\begin{eqnarray}
 a=2\sinh\epsilon_0\,,\qquad b=\frac{\cosh(2\epsilon_0+\sigma_0)}{\cosh(\epsilon_0+\sigma_0)}\,,\qquad
c=\frac{\sinh\sigma_0}{\cosh(\epsilon_0+\sigma_0)}\,,\qquad
b_0=e^{\epsilon_0}\,,\qquad c_0=e^{-\epsilon_0}\,.
\end{eqnarray}
Applying the formula (\ref{sum_math}) and subsequently expanding for small $\epsilon_0$, we get
\begin{eqnarray}
\sum_{E=1}^\Lambda \left(\Omega^+_{E+\frac{1}{2}}+\Omega^+_{-E+\frac{1}{2}}\right)&\simeq&
\frac{\tanh\sigma_0-1}{2}\left(\ln\Lambda+\ln(2\epsilon_0)\right)-\ln\Gamma(1+\tanh\sigma_0)\,.\cr &&
\end{eqnarray}
Notice that in the limit $\sigma_0\to\infty$ this expression vanishes, as it should. The last term in the sum (\ref{sumintferm}) can be evaluated using (\ref{secform}) and gives
\begin{eqnarray}
-\frac{\mu}{2}\sum_{E=1}^\infty e^{-\mu E}\left(\Omega^+_{E+\frac{1}{2}}-\Omega^+_{-E+\frac{1}{2}}\right)
&=&-\ln\sqrt{\frac{1+\tanh\sigma_0}{2}}\,.
\end{eqnarray}
Putting everything together, we get
\begin{eqnarray}
	\frac{1}{2}\sum_{E\,\in\,\mathds{Z}+\frac{1}{2}}e^{-\mu|E|}\left(\Omega^+_E+\Omega^+_{-E}\right)
	&=&  \frac{\tanh\sigma_0-1}{2}\left(\ln\Lambda+\ln(2\epsilon_0)\right)-\ln\Gamma(1+\tanh\sigma_0)\,.
\end{eqnarray}
The $\alpha=-1$ case is identical and the total contribution from the fermions becomes
\begin{eqnarray}
-\frac{4}{2}\times 2\Big(
\frac{\tanh\sigma_0-1}{2}\left(\ln\Lambda+\ln(2\epsilon_0)\right)-\ln\Gamma(1+\tanh\sigma_0)
\Big)\,.
\end{eqnarray}
Notice that, with this order of limits, not only we get a surviving logarithmic divergence in $\epsilon_0$, but the logarithmic divergence in $\Lambda$ does not cancel against the similar divergence in the bosonic sector because of the extra factor of $1/2$.


\bibliographystyle{JHEP}
\bibliography{WLoops-bib}
\end{document}